\documentclass[a4paper,11pt]{article}
\newcommand{\s}[1]{{\huge\textsf{\textbf{#1}}}}
\usepackage[left=20mm,right=20mm,top=20mm,bottom=20mm]{geometry}
\usepackage{amsmath}
\usepackage{amsfonts}
\usepackage{amssymb}
\usepackage{amsbsy}
\usepackage{color,graphics}
\usepackage{setspace}
\usepackage{epsfig}
\usepackage{subfigure}
\usepackage{cite}
\usepackage{graphicx}
\usepackage{latexsym,multicol}
\usepackage{tcolorbox}

\graphicspath{{./LCE-plates-figs/}}
\begin{document}
\title{\s{A plate theory for nematic liquid crystalline solids}}
\author{L. Angela Mihai\footnote{School of Mathematics, Cardiff University, Senghennydd Road, Cardiff, CF24 4AG, UK, Email: \texttt{MihaiLA@cardiff.ac.uk}}
	\qquad Alain Goriely\footnote{Mathematical Institute, University of Oxford, Woodstock Road, Oxford, OX2 6GG, UK, Email: \texttt{goriely@maths.ox.ac.uk}}
}	
\maketitle

\begin{abstract}
\hrule\vskip 12pt
We derive a F\"{o}ppl-von K\'{a}rm\'{a}n-type constitutive model for solid liquid crystalline plates where the nematic director may or may not rotate freely relative to the elastic network. To obtain the reduced two-dimensional model, we rely on the deformation decomposition of a nematic solid into an elastic deformation and a natural shape change. The full solution to the resulting equilibrium equations consists of both the deformation displacement and stress fields. The model equations are applicable to a wide range of thin nematic bodies subject to optothermal stimuli and mechanical loads. For illustration, we consider certain reversible natural shape changes in simple systems which are stress free, and their counterparts, where the natural deformations are blocked and internal stresses appear. More general problems can be addressed within the same framework.
\vskip 6pt
\hrule\vskip 6pt
\noindent{\bf Keywords:} liquid crystals, nematic solids, elastic plates, nonlinear deformation, multiplicative decomposition, stress.
\end{abstract}


\section{Introduction}

Liquid crystalline (LC) solids are complex materials that combine the elasticity of polymeric solids with the self-organisation of liquid crystal structures \cite{deGennes:1975,Finkelmann:1981:FKR}. Due to their molecular architecture, consisting of cross-linked networks of polymeric chains containing liquid crystal mesogens, their deformations are typically large and nonlinear, and can arise spontaneously and reversibly under certain external stimuli (heat, light, solvents, electric or magnetic field) \cite{Agrawal:2014:etal,Brommel:2013:etal,Corbett:2008:CW,Davidson:2019:etal,Finkelmann:2001:FNPW,Kaiser:2009:etal,Kupfer:1991:KF,Kupfer:1994:KF,Torras:2011:etal,Uruyama:2005:UHT,Uruyama:2006:UHT,Wang:2017:WTHC,Ware:2015:etal,Wei:2012:WY,Winkler:2010:etal,Zentel:1986,Zhao:2020:ZL}. These qualities suggest many avenues for technological applications, but more research efforts are needed before they can be exploited on an industrial scale  \cite{Camacho:2004:etal,DeSimone:2015:DSGN,Ford:2019:etal,Gelebart:2017:etal,Haghiashtiani:2018:etal,Mori:2020:etal,Prevot:2018:etal,Tian:2018:etal,Tottori:2012:etal,Ula:2018:etal,vanOosten:2007:OHBB,Wan:2018.etal,Wang:2019:etal,Wie:2016:WSW}.

For thin nematic bodies, large deformations have been studied extensively, both theoretically and in laboratory. Rectangular geometries were assumed in \cite{Aharoni:2014:ASK,Cirak:2014:CLBW}, circular discs were considered in
\cite{Ahn:2015:ALC,Ahn:2016:etal,Konya:2016:KGPS,Kowalski:2017:etal,Modes:2010:MBW,Modes:2011:MBW,Modes:2011:MW,Mostajeran:2015,Mostajeran:2016:MWWW,Pismen:2014,Warner:2018:WM}, thin ribbons were treated theoretically in \cite{Agostiniani:2017a:AdS,Agostiniani:2017b:AdS,Agostiniani:2017:AdSK,Tomassetti:2017:TV} and experimentally in \cite{Sawa:2011:etal,Sawa:2013:etal,Uruyama:2013}. Different molecular structures and compositions, and the complex morphing behaviour that can be achieved in liquid crystal polymer networks were reviewed in \cite{deHaan:2014:etal,Kuenstler:2019:KH,Modes:2016:MW,Pang:2019:etal,Warner:2020,White:2015:WB}. 

At the constitutive level, for ideal monodomain nematic solids, where the mesogens are aligned throughout the material, a general formulation is usually provided by the phenomenological neoclassical strain-energy function proposed in \cite{Bladon:1994:BTW,Warner:1988:WGV,Warner:1991:WW}. This model is based on the molecular network theory of rubber elasticity\cite{Treloar:2005}, where the parameters are directly measurable experimentally or derived from macroscopic shape changes \cite{Warner:1996:WT,Warner:2007:WT}. For nematic polydomains, where the mesogens are separated into many domains, such that in every domain, they that are aligned along a local director, in \cite{Biggins:2009:BW,Biggins:2012:BWB}, it is assumed that each domain has the same strain-energy density as a monodomain. Extensions to strain-energy functions for nematic elastomer plates with out-of-plane heterogeneities were proposed in \cite{Agostiniani:2017a:AdS,Agostiniani:2017b:AdS,Agostiniani:2017:AdSK}. General continuum mechanical theories for nematic elastomers are provided in \cite{Anderson:1999:ACF,Zhang:2019:etal}. 

In this study, we derive a reduced two-dimensional (2D) model describing the equilibrium of thin nematic solids made of a liquid crystalline material subject to optothermal stimuli and mechanical loads. We first define the strain-energy function for a solid nematic material where the director may or may not rotate freely relative to the elastic matrix (Section~\ref{NLC:sec:model}), then formulate the corresponding stress tensors similar to those from the finite elasticity theory (Section~\ref{NLC:sec:stress}). Here, we take as reference configuration the isotropic phase at high temperature \cite{Cirak:2014:CLBW,DeSimone:1999,DeSimone:2000:dSD,DeSimone:2002:dSD,DeSimone:2009:dST} (inspired by the classical work of Flory (1961) \cite{Flory:1961} on polymer elasticity), rather than the nematic phase in which the cross-linking was produced \cite{Anderson:1999:ACF,Bladon:1994:BTW,Verwey:1996:VWT,Warner:1994:WBT,Warner:1988:WGV,Warner:1991:WW,Zhang:2019:etal}. The relation between the trace formula of \cite{Bladon:1994:BTW} (using as reference orientation the one corresponding to the cross-linking state) and the neo-Hookean-based strain-energy density defined in \cite{DeSimone:1999} (with a ``virtual'' isotropic state as reference configuration) is explained in \cite{DeSimone:2009:dST}. Similar nematic strain-energy densities based on other classical hyperelastic models (e.g., Money-Rivlin, Ogden) are also discussed in \cite{DeSimone:2009:dST}. In adopting the isotropic phase as reference configuration, we follow Cirak et al. (2014) \cite{Cirak:2014:CLBW}, where strain-energy functions with either free or frozen nematic director are defined, and the director has an initial direction which may be spatially varying. Our choice is phenomenologically motivated by the multiplicative decomposition of the deformation gradient from the reference configuration to the current configuration into an elastic distortion followed by a natural (stress free) shape change. This multiplicative decomposition is similar to those found in the constitutive theories of thermoelasticity, elastoplasticity, and growth \cite{goriely17,Lubarda:2004}, but it is fundamentally different as the stress free geometrical change of liquid crystalline solids is superposed on the elastic deformation, which is applied directly to the reference state. Such difference is important since, although the elastic configuration obtained by this deformation may not be observed in practice, it may still be possible for the nematic body to assume such a configuration under suitable external stimuli. The elastic stresses can then be used to analyse the final deformation where the particular geometry also plays a role. However, in liquid crystalline materials, asymmetric Cauchy stresses generally occur, unlike in purely elastic materials \cite[p.~80]{Warner:2007:WT}. We employ the method of asymptotic expansions, with the thickness of the body as the small parameter, and show that the leading term of the expansion is the solution of a system of equations of the F\"{o}ppl-von K\'{a}rm\'{a}n-type \cite{Foppl:1907,vonKarman:1910} (Section~\ref{NLC:sec:plate}). A similar model for the elastic growth of thin biological plates was developed in \cite{Dervaux:2009:DCBA} (see also \cite{Dervaux:2008:DBA}). For an initial application of the plate theory, we consider `spontaneous' nonlinear deformations of annular circular discs (Section~\ref{NLC:sec:rings}). We conclude with a summary of these results and further remarks (Section~\ref{NLC:sec:conclude}).

\section{An ideal nematic solid}\label{NLC:sec:model}
\begin{figure}[ht]
	\begin{center}
		\includegraphics[width=0.6\textwidth]{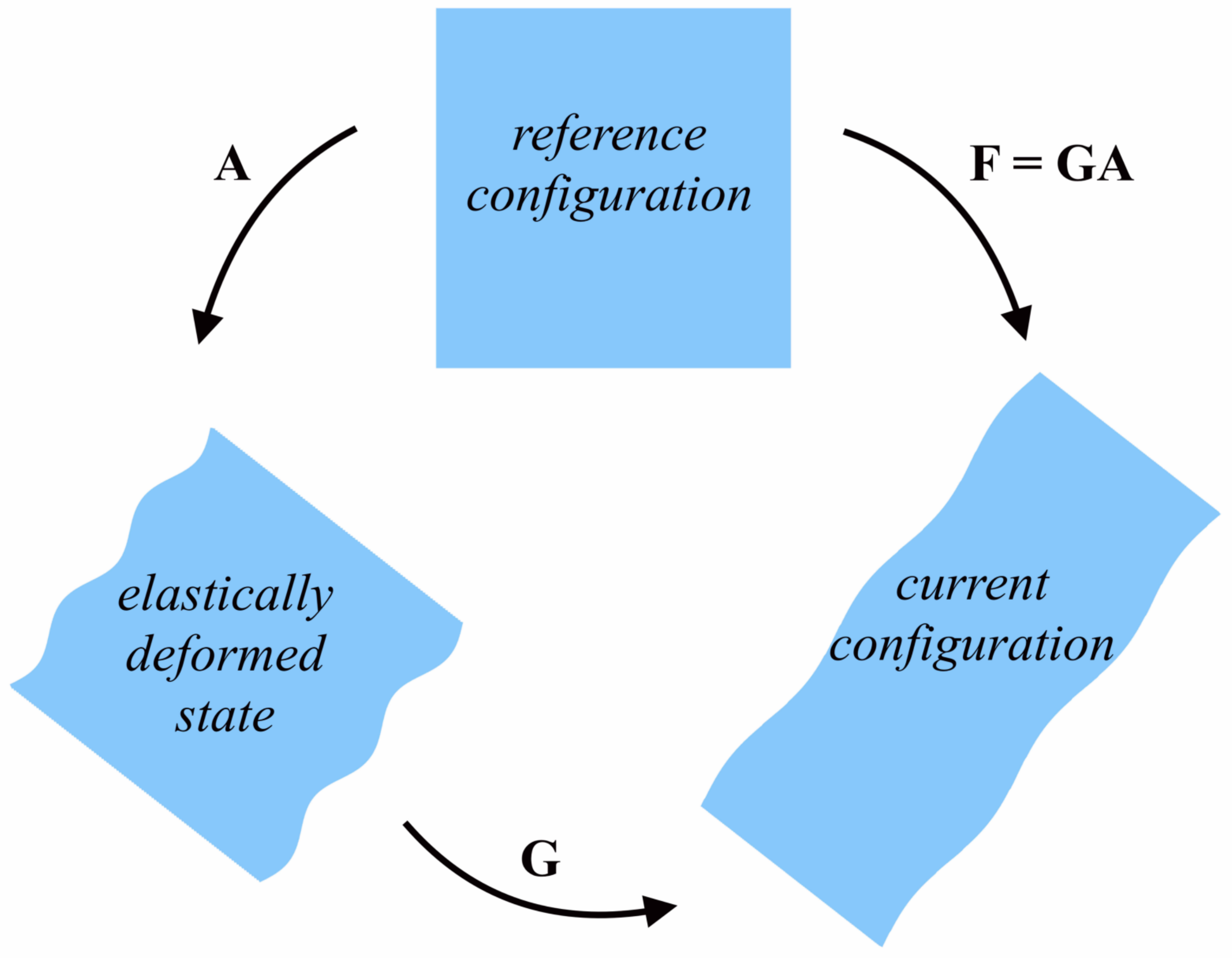}
	\end{center}
	\caption{Schematic of composite deformation of a nematic solid from a reference configuration to the current configuration, by a natural shape change superposed on an elastic distortion.}\label{NLC:fig:FGA}
\end{figure}

For an ideal nematic liquid crystalline (NLC) solid, the neoclassical strain-energy density function takes the generic form
\begin{equation}\label{NLC:eq:Wnc}
W^{(nc)}(\textbf{F},\textbf{n})=W(\textbf{A}),
\end{equation}
where $\textbf{F}$ represents the deformation gradient from the isotropic state, $\textbf{n}$ denotes the unit vector (or `director') for the orientation of the nematic field, and $W(\textbf{A})$ is the strain-energy density of the isotropic polymer network, depending only on the (local) elastic deformation tensor $\textbf{A}$. The tensors $\textbf{F}$ and $\textbf{A}$ satisfy the relation (see Figure~\ref{NLC:fig:FGA})
\begin{equation}\label{NLC:eq:FGA}
\textbf{F}=\textbf{G}\textbf{A},
\end{equation}
where $\textbf{G}$ is the `natural' (or `spontaneous') deformation tensor defining a change of frame of reference from the isotropic phase to a nematic phase, and is equal to
\begin{equation}\label{NLC:eq:G:nu}
\textbf{G}=a^{1/3}\textbf{n}\otimes\textbf{n}+a^{-\nu/3}\left(\textbf{I}-\textbf{n}\otimes\textbf{n}\right)=a^{-\nu/3}\textbf{I}+\left(a^{1/3}-a^{-\nu/3}\right)\textbf{n}\otimes\textbf{n}.
\end{equation}
In \eqref{NLC:eq:G:nu}, $a>0$ is a temperature-dependent stretch parameter, $\nu$ denotes the optothermal analogue to the Poisson ratio \cite{Modes:2011:MW}, relating responses in directions parallel or perpendicular to the director $\textbf{n}$, $\otimes$ stands for the tensor product of two vectors, and $\textbf{I}=\text{diag}(1,1,1)$ is the identity tensor. For example, spontaneous extensions or contractions of magnitude 10\%-400\% along $\textbf{n}$ can occur in LC solids, while elastomers have $\nu=1/2$, i.e., their volume remains unchanged during deformation, and glasses have $\nu\in(1/2,2)$ \cite{vanOosten:2007:OHBB,Warner:2020}. We assume here that $a$ and $\nu$ are spatially-independent (no differential swelling). These parameters can be estimated independently (e.g., by examining the thermal or light-induced response of the nematic elastomer with uniform planar alignment \cite{Kowalski:2017:etal}). The ratio $r=a^{1/3}/a^{-\nu/3}=a^{(\nu+1)/3}$ represents the anisotropy parameter, which, in an ideal nematic solid, is the same in all directions. In the nematic phase, both the cases with $r>1$ (prolate molecules) and $r<1$ (oblate molecules) are possible; when $r=1$, the energy function reduces to that of an isotropic hyperelastic material \cite{deHaan:2014:etal}. Note that $\textbf{G}$ given by \eqref{NLC:eq:G:nu} is symmetric, i.e., $\textbf{G}=\textbf{G}^{T}$ (with the superscript ``$T$'' denoting the transpose operation), whereas the elastic tensor $\textbf{A}$ may not be symmetric in general.

The hyperelastic strain-energy function $W(\textbf{A})$ in \eqref{NLC:eq:Wnc} is minimised by any deformation satisfying $\textbf{A}\textbf{A}^{T}=\textbf{I}$ \cite{Ogden:1997,TruesdellNoll:2004}, while the corresponding nematic strain-energy
$W^{(nc)}(\textbf{F},\textbf{n})$ is minimised by any deformation satisfying $\textbf{F}\textbf{F}^{T}=\textbf{G}^2$. Hence, every pair $(\textbf{G}\textbf{R},\textbf{n})$, where $\textbf{R}$ is an arbitrary rigid-body rotation (i.e., $\textbf{R}^{-1}=\textbf{R}^{T}$ and $\det\textbf{R}=1$), is a natural (i.e., stress free) state for this material model.

In NLC solids, the director $\textbf{n}$ is an observable (spatial) quantity. Denoting by $\textbf{n}_{0}$ the reference orientation of the local director corresponding to the cross-linking state, $\textbf{n}$ may differ from $\textbf{n}_{0}$ both by a rotation and a change in $r$. For NLC elastomers, which are weakly cross-linked, the nematic director can rotate freely, and the material exhibits isotropic mechanical effects. In NLC glasses, which are densely cross-linked, the nematic director $\textbf{n}$ cannot rotate relative to the elastic matrix, but changes through convection due to elastic strain, and satisfies \cite{Cirak:2014:CLBW,Modes:2010:MBW,Modes:2011:MBW,Modes:2011:MW,Mostajeran:2015}
\begin{equation}\label{NLC:eq:n0n}
\textbf{n}=\frac{\textbf{F}\textbf{n}_{0}}{|\textbf{F}\textbf{n}_{0}|}.
\end{equation}
This constraint enables patterning of the director field at cross-linking and guarantees that the ``written-in'' pattern remains virtually the same during natural shape changes \cite{Modes:2010:MBW,Modes:2011:MBW,Warner:2010:WMC}. The elastic anisotropy of NLC materials where the director cannot rotate was investigated experimentally in \cite{Finkelmann:2001:FGW,Mistry:2019:MG}. Natural strains in NLC glasses are typically of up to 4\%,  whereas for NLC elastomers, these may be up to 400\%.

\section{Stress tensors}\label{NLC:sec:stress}

We restrict our attention to the case when $W(\textbf{A})$ in \eqref{NLC:eq:Wnc} describes an incompressible neo-Hookean material \cite{Treloar:1944}, i.e.,
\begin{equation}\label{NLC:eq:W:NH}
W(\textbf{A})=\frac{\mu}{2}\left[\text{tr}\left(\textbf{A}\textbf{A}^{T}\right)-3\right],
\end{equation}
where ``tr'' denotes the trace operator, and $\mu>0$ represents the constant shear modulus at small strain. 

Of particular significance are the left and right Cauchy-Green tensors defined, respectively, by
\begin{equation}\label{NLC:eq:BC}
\textbf{B}=\textbf{A}\textbf{A}^{T}
\qquad\mbox{and}\qquad
\textbf{C}=\textbf{A}^{T}\textbf{A}.
\end{equation}
Using these deformation tensors, the elastic Almansi strain tensor is equal to \cite[pp.~90-91]{Ogden:1997}
\begin{equation}\label{NLC:eq:Almansi}
\textbf{e}=\frac{1}{2}\left(\textbf{I}-\textbf{B}^{-1}\right),
\end{equation}
and the elastic Green-Lagrange strain tensor is \cite[pp.~89-90]{Ogden:1997}
\begin{equation}\label{NLC:eq:GL}
\textbf{E}=\frac{1}{2}\left(\textbf{C}-\textbf{I}\right).
\end{equation}

For the given hyperelastic material, the Cauchy (true) stress tensor (representing the internal force per unit of deformed area acting within the deformed solid) takes the form
\begin{equation}\label{NLC:eq:cauchy:iso}
\textbf{T}=\left(\det\textbf{A}\right)^{-1}\frac{\partial W}{\partial\textbf{A}}\textbf{A}^{T}-p\textbf{I}=\mu\textbf{B}-p\textbf{I},
\end{equation}
where $p$ denotes the Lagrange multiplier for the internal constraint $\det\textbf{A}=1$ \cite[pp.~198-201]{Ogden:1997}. The Cauchy stress tensor $\textbf{T}$ defined by \eqref{NLC:eq:cauchy:iso} is symmetric and coaxial (i.e., it has the same eigenvectors) with the left Cauchy-Green tensor $\textbf{B}$ given by \eqref{NLC:eq:BC}, and with the Almansi strain tensor $\textbf{e}$ given by \eqref{NLC:eq:Almansi}.

The corresponding first Piola-Kirchhoff stress tensor (representing the internal force per unit of undeformed area acting within the deformed solid) is equal to
\begin{equation}\label{NLC:eq:1PK:iso}
\textbf{P}=\textbf{T}\text{Cof}(\textbf{A})=\frac{\partial W}{\partial\textbf{A}}-p\textbf{A}^{-T}=\mu\textbf{A}-p\textbf{A}^{-T},
\end{equation}
where $\text{Cof}(\textbf{A})=\left(\det\textbf{A}\right)\textbf{A}^{-T}$ is the cofactor of $\textbf{A}$. This stress tensor is not symmetric in general. Its transpose $\textbf{P}^{T}$ is known as the nominal (engineering) stress tensor \cite[pp.~152-153]{Ogden:1997}.

The associated second Piola-Kirchhoff stress tensor is
\begin{equation}\label{NLC:eq:2PK:iso}
\textbf{S}=\textbf{A}^{-1}\textbf{P}=2\frac{\partial W}{\partial\textbf{C}}-p \textbf{C}^{-1}=\mu\textbf{I}-p \textbf{C}^{-1},
\end{equation}
and this is symmetric and coaxial with the right Cauchy-Green tensor $\textbf{C}$ defined by \eqref{NLC:eq:BC} and with the Green-Lagrange strain tensor $\textbf{E}$ defined by \eqref{NLC:eq:GL}.

In the small strain limit, the Cauchy stress tensor $\textbf{T}$ and the Piola-Kirchhoff stress tensors $\textbf{P}$ and $\textbf{S}$ coincide \cite{Mihai:2017:MG}.

Based on the hyperelastic model defined by \eqref{NLC:eq:W:NH}, we construct the following strain-energy function of the form \eqref{NLC:eq:Wnc} corresponding to an ideal nematic solid \cite{Cirak:2014:CLBW,DeSimone:2000:dSD,DeSimone:2009:dST},
\begin{equation}\label{NLC:eq:Wnc:NH}
W^{(nc)}(\textbf{F},\textbf{n})=\frac{\mu}{2}\left\{a^{2\nu/3}\left[\text{tr}\left(\textbf{F}\textbf{F}^{T}\right)-\left(1- a^{-2(1+\nu)/3}\right)\textbf{n}\cdot\textbf{F}\textbf{F}^{T}\textbf{n}\right]-3\right\}.
\end{equation}
Despite its dependence on $\mathbf{n}$, this strain-energy function is isotropic \cite{DeSimone:2009:dST,Finkelmann:2001:FGW}, and it can therefore be expressed equivalently in terms of the principal stretch ratios \cite{Cirak:2014:CLBW}. To see this, we distinguish two cases: first, when the nematic director is `free' to rotate relative to the elastic matrix, in which case it tends to align in the direction of the maximum principal stretch ratio, and second, when the nematic director is `frozen' and satisfies condition \eqref{NLC:eq:n0n}. For both these cases, in order to construct the corresponding plate models, we are interested in the stress tensors of the deformed nematic material in terms of the stresses in the base polymeric network. The relations between these stress tensors are presented in Appendix~\ref{NLC:sec:append:stress}, where the superscript `$(nc)$' is used when denoting stresses in the nematic material.

\section{The 3D equilibrium equations}\label{NLC:sec:equilibrium}

We consider a solid nematic body characterised by the strain-energy function defined by \eqref{NLC:eq:Wnc:NH}, and occupying a compact domain $\bar\Omega\subset\mathbb{R}^{3}$, such that the interior of the body is an open, bounded, connected set $\Omega\subset\mathbb{R}^{3}$, and its boundary $\partial\Omega=\bar\Omega\setminus\Omega$ is Lipschitz continuous (in particular, we assume that a unit normal vector $\textbf{N}$ exists almost everywhere on $\partial\Omega$). The elastic energy stored by the body is equal to \cite[p.~205]{Ogden:1997}
\begin{equation}\label{NLC:eq:Enc}
E=\int_{\Omega}\left[W^{(nc)}\left(\textbf{F}\right)-p^{(nc)}\left(\det\textbf{F}-\det\textbf{G}\right)\right]dV,
\end{equation}
where $p^{(nc)}\left(\det\textbf{F}-\det\textbf{G}\right)$ enforces the condition that $\det\textbf{F}=\det\textbf{G}$, with $\det\textbf{G}$ representing the local change of volume due to the `spontaneous' deformation.

Assuming that, on part of its boundary, $\Gamma_{N}\subset\partial\Omega$, the body is subject to a surface force (traction) $\textbf{f}_{N}$ that is also derivable from a potential function, the energy minimisation problem in the absence of body forces is:
\begin{equation}
\mbox{Minimise} \qquad E_{t}=E- \int_{\Gamma_{N}}\textbf{f}_{N}\cdot\textbf{u}\,dA\qquad \mbox{subject to}\qquad \det\textbf{F}=\det\textbf{G},
\end{equation}
where $E_{t}$ represents the \emph{total potential energy} over all displacement fields $\textbf{u}(\textbf{X})=\textbf{x}-\textbf{X}$, with gradient tensor $\nabla\textbf{u}=\textbf{F}-\textbf{I}$, such that the kinematic constraint \eqref{NLC:eq:FGA} holds, and the last integral represents the work done by the external force. Following a standard procedure of computing variations $\delta E$ in $E$ for
infinitesimal variations $\delta\textbf{u}$ in $\textbf{u}$, the following first variation in $E_{t}$ is obtained (see \cite[p.~310-312]{Ogden:1997}),
\begin{equation}
\begin{split}
\delta E_{t}&=\delta E-\int_{\Gamma_{N}}\textbf{f}_{N}\cdot\delta\textbf{u}\,dA\\
&=-\int_{\Omega}\textbf{P}^{(nc)}:\delta\textbf{F}dV- \int_{\Gamma_{N}}\textbf{f}_{N}\cdot\delta\textbf{u}\,dA\\
&=\int_{\Omega}\text{Div}\ \textbf{P}^{(nc)}\cdot\delta\textbf{u}\,dV,
\end{split}
\end{equation}
where $\textbf{P}^{(nc)}$ is the first Piola-Kirchhoff stress tensor (given explicitly by  \eqref{NLC:eq:1PK:nc} in Appendix A).

We arrive at \emph{the principle of stationary potential energy} stating that: of all admissible displacements fields that the body can assume, those for which the total potential energy assumes a stationary value, such that $\delta E_{t}=0$, correspond to static equilibrium state described by
\begin{equation}\label{NLC:eq:balance}
-\text{Div}\ \textbf{P}^{(nc)}(\textbf{X})=\textbf{0},
\end{equation}
together with the traction boundary condition causing the deformation, $\textbf{P}^{(nc)}\textbf{N}=\textbf{f}_{N}$, where $\textbf{N}$ is the outward unit normal vector to the external surface $\Gamma_{N}$.

To simplify the mechanical analysis, it is convenient to rewrite the above equations equivalently in terms of the elastic stresses for the base polymeric network. Since the elastic deformation is applied directly to the reference state, using the relations between the stress tensors in the nematic material presented in Appendix~\ref{NLC:sec:append:stress}, we obtain
\begin{equation}
\begin{split}
\delta E
&=-\int_{\Omega}\textbf{P}^{(nc)}:\delta\textbf{F}dV\\
&=-\int_{\Omega}\textbf{G}^{-1}\textbf{P}:\delta\left(\textbf{G}\textbf{A}\right)dV\\
&=-\int_{\Omega}\textbf{G}^{-1}\textbf{A}\textbf{S}:\delta\left(\textbf{G}\textbf{A}\right)dV\\
&=-\int_{\Omega}\textbf{S}:\left[\left(\textbf{G}^{-1}\textbf{A}\right)^{T}\delta\left(\textbf{G}\textbf{A}\right)\right] dV
\end{split}
\end{equation}
where $\textbf{P}$  and $\textbf{S}$ are the Piola-Kirchhoff stress tensors given by \eqref{NLC:eq:1PK:iso} and \eqref{NLC:eq:2PK:iso}, respectively. Hence,
\begin{equation}\label{NLC:eq:deltaE}
\delta E=-\int_{\Omega}\textbf{S}:\delta\textbf{E}\, \, dV,
\end{equation}
where $\textbf{E}$ is the Green-Lagrange strain tensor defined by \eqref{NLC:eq:GL}. The equilibrium equation \eqref{NLC:eq:balance} is then equivalent to
\begin{equation}\label{NLC:eq:balance:iso}
-\text{Div}\ \textbf{S}(\textbf{X})=\textbf{0},
\end{equation}
with the corresponding boundary condition $\textbf{S}\textbf{N}=\textbf{A}^{-1}\textbf{G}\textbf{f}_{N}$.

The potential energy \eqref{NLC:eq:Enc}  can be expressed equivalently as
\begin{equation}\label{NLC:eq:E}
E=\int_{\Omega}\left[W\left(\textbf{A}\right)-p\left(\det\textbf{A}-1\right)\right]dV,
\end{equation}
where $W(\textbf{A})$ is the elastic strain-energy function on which the nematic strain-energy function $W^{(nc)}(\textbf{A})$ is based, and $p\left(\det\textbf{A}-1\right)$ enforces the incompressibility condition $\det\textbf{A}=1$. 

\section{The nematic plate model}\label{NLC:sec:plate}

Our goal is to devise a 2D model for a nematic solid which is sufficiently thin so that it can be approximated by a plate equation. In our derivation of the constitutive equations for the nematic plate model, we rely on the following conditions assumed a priori:
\begin{itemize}
	\item[(P1)] Surface normals to the plane of the plate remain perpendicular to the plate after deformation;
	\item[(P2)] Changes in the thickness of the plate during deformation are negligible;
	\item[(P3)] The stress field in the deformed plate is parallel to the mid-surface.
	\end{itemize}
These kinematic assumptions from the classical theory of plates (the Kirchhoff-Love hypotheses) \cite{Love:1888} are valid also for the F\"{o}ppl-von K\'{a}rm\'{a}n theory \cite{Foppl:1907,vonKarman:1910} (see also \cite{Ciarlet:1980,Erbay:1997,Landau:1986:LL}). They are appropriate for nematic solids which are typically prepared as thin sheets, so that suitable heat transfer can be ensured \cite{deHaan:2014:etal,Kuenstler:2019:KH,Modes:2016:MW,Pang:2019:etal,Ula:2018:etal,Warner:2020,White:2015:WB}, and allow to reduce the dimensionality of the problem, which is useful computationally.

\begin{figure}[ht]
	\begin{center}
		\includegraphics[width=0.65\textwidth]{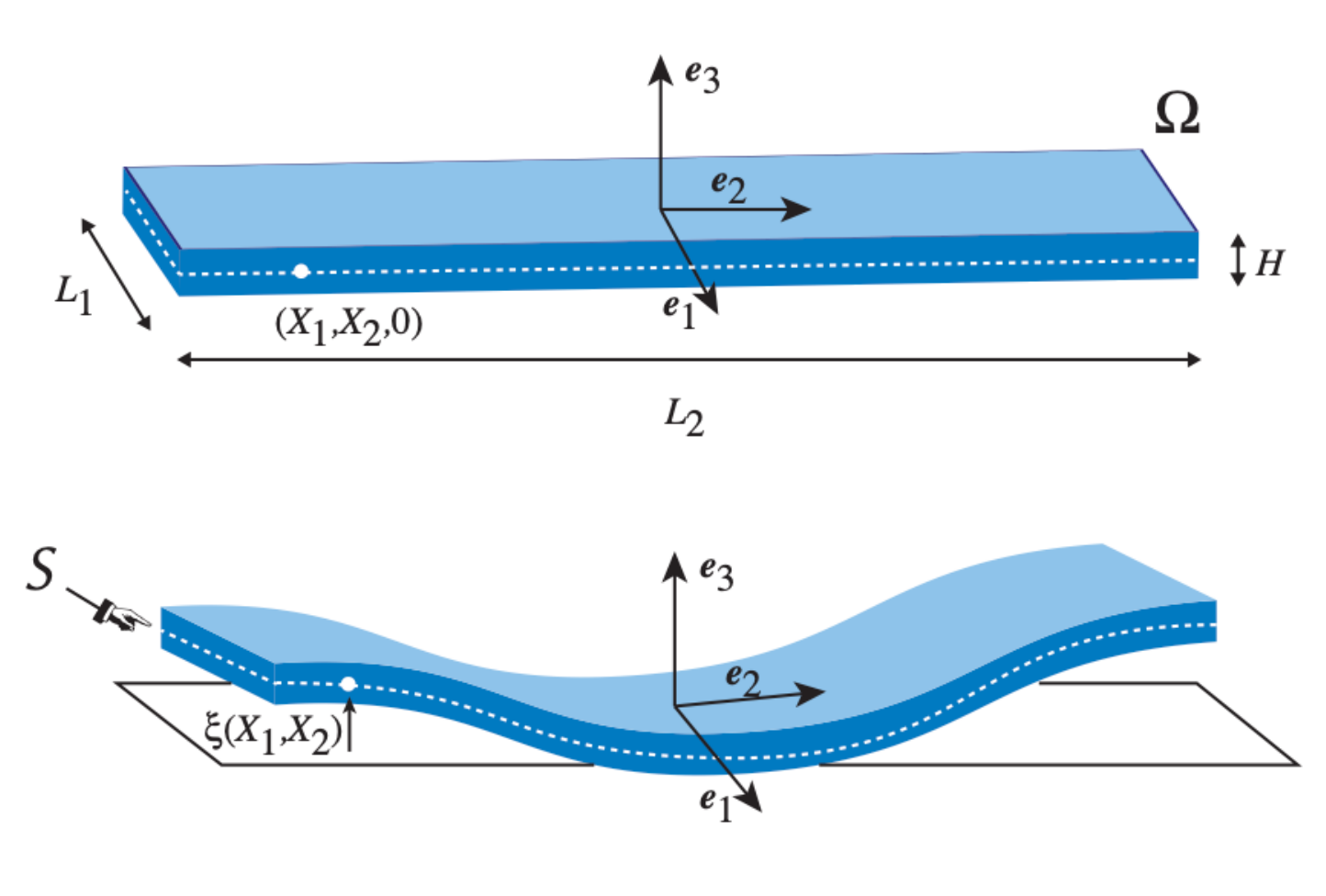}
	\end{center}
	\caption{Geometry of the F\"oppl-von K\'arm\'an plate: undeformed (reference) configuration (top), and deformed configuration (bottom). The deformation is fully characterised by  the vertical displacement $\xi(X_{1},X_{2})$ of the mid-surface $\mathsf{S}=(-L_{1}/2,L_{1}/2)\times(-L_{2}/2,L_{2}/2)\times\{0\}$.}
	\label{NLC:fig:Fig-plates}
\end{figure}

Specifically, we assume that $\Omega=(-L_{1}/2,L_{1}/2)\times(-L_{2}/2,L_{2}/2)\times(-H/2,H/2)$ is the domain occupied by the thin nematic body in the undeformed state, where $H$ is small compared to $L_{1}$ and $L_{2}$, which are both of order $L$, i.e., $H\ll L$. We denote by $\mathsf{S}=(-L_{1}/2,L_{1}/2)\times(-L_{2}/2,L_{2}/2)\times\{0\}$ the mid-surface. The nematic plate in the undeformed state is illustrated in Figure~\ref{NLC:fig:Fig-plates}-top. In a Cartesian system of coordinates, $\textbf{X}=\left(X_{1},X_{2},X_{3}\right)\in\Omega$ is given by $\textbf{X}=X_{1}\textbf{e}_{1}+X_{2}\textbf{e}_{2}+X_{3}\textbf{e}_{3}$. 

\subsection{Free nematic director}\label{NLC:sec:nfree}

We consider in detail the case when $\textbf{F}$ and $\textbf{n}$ are independent variables. First, we express the elastic strain-energy function described by \eqref{NLC:eq:W:NH} in the equivalent form
\begin{equation}\label{NLC:eq:W:NH:I1}
\mathcal{W}(I_{1})=\frac{\mu}{2}\left(I_{1}-3\right),
\end{equation}
where $I_{1}=\text{tr}(\textbf{B})=\text{tr}(\textbf{C})$ is the first principal invariant of the Cauchy-Green tensors given by \eqref{NLC:eq:BC}. Taking into account that the body is thin, we can approximate this function as follows,
\begin{equation}\label{NLC:eq:W:app}
\mathcal{W}(I_{1})\approx \mathcal{W}_{0}\left(I_{1}^{(0)}\right)+\mathcal{W}_{1}\left(I_{1}^{(1)}\right)X_{3}+\mathcal{W}_{2}\left(I_{1}^{(2)}\right)X_{3}^2,
\end{equation}
with
\begin{equation}
\mathcal{W}_{0}\left(I_{1}^{(0)}\right)=\frac{\mu}{2}\left(I_{1}^{(0)}-3\right),\qquad
\mathcal{W}_{1}\left(I_{1}^{(1)}\right)=\frac{\mu}{2}I_{1}^{(1)}\qquad
\mathcal{W}_{2}\left(I_{1}^{(2)}\right)=\frac{\mu}{2}I_{1}^{(2)},
\end{equation}
and
\begin{equation}\label{NLC:eq:I1app}
I_{1}\approx I_{1}^{(0)}+I_{1}^{(1)}X_{3}+I_{1}^{(2)}X_{3}^2,
\end{equation}
where ``$\approx$'' is used to denote the Taylor expansion up to second order in $X_{3}$.

The corresponding  expansions for the determinant of the elastic deformation and the hydrostatic pressure are, respectively,
\begin{equation}\label{NLC:eq:detAapp}
\det \textbf{A}\approx \mathcal{D}^{(0)}+\mathcal{D}^{(1)}X_{3}+\mathcal{D}^{(2)}X_{3}^2,
\end{equation}
\begin{equation}\label{NLC:eq:p0app}
p\approx p^{(0)}+p^{(1)}X_{3}+p^{(2)}X_{3}^2.
\end{equation}
We will show that both $p^{(1)}$ and $p^{(2)}$ vanish and, therefore, we do not include them explicitly in the subsequent expressions.

For plates, when large out-of-plane deflections occur, the bending generally involves stretching. In this case, the energy defined by \eqref{NLC:eq:E} can be approximated as follows,
\begin{equation}\label{NLC:eq:Eapp}
E\approx E_{stretch}+E_{bend},
\end{equation}
where $E_{stretch}$ and $E_{bend}$ represent the stretching and bending contributions, respectively. Explicitly, they read
\begin{equation}\label{NLC:eq:Estretch}
E_{stretch}=H\int_{-L_{2}/2}^{L_{2}/2}\int_{-L_{1}/2}^{L_{1}/2}\left[\mathcal{W}_{0}\left(I_{1}^{(0)}\right)-p^{(0)}\left(\mathcal{D}^{(0)}-1\right)\right]dX_{1}dX_{2}
\end{equation}
and
\begin{equation}\label{NLC:eq:Ebend}
E_{bend}=\frac{H^3}{12}\int_{-L_{2}/2}^{L_{2}/2}\int_{-L_{1}/2}^{L_{1}/2}\left[\mathcal{W}_{2}\left(I_{1}^{(2)}\right)-p^{(0)} \mathcal{D}^{(2)}\right]dX_{1}dX_{2}.
\end{equation}
 Note that, due to the symmetry of the domain relative to the mid-surface $\mathsf{S}$, the integral of the first-order term in the approximate strain-energy function given by \eqref{NLC:eq:W:app} vanishes identically.

\subsubsection{Model reduction}

We denote by $\textbf{D}=\nabla\textbf{u}=\textbf{F}-\textbf{I}$ the gradient tensor of the displacement field. For the thin nematic body, we write the displacement vector as
\begin{equation}
\textbf{u}=
\left[
\begin{array}{c}
u_{1}(X_{1},X_{2},X_{3})\\
u_{2}(X_{1},X_{2},X_{3})\\
u_{3}(X_{1},X_{2},X_{3})
\end{array}
\right]
=\left[
\begin{array}{c}
u_{1}(X_{1},X_{2},X_{3})\\
u_{2}(X_{1},X_{2},X_{3})\\
\xi(X_{1},X_{2})+w(X_{1},X_{2},X_{3})
\end{array}
\right],
\end{equation}
such that:
\begin{equation}
\begin{split}
u_{1}(X_{1},X_{2},X_{3})&\approx u_{1}^{(0)}(X_{1},X_{2})+X_{3}u_{1}^{(1)}(X_{1},X_{2})+\cdots,\\
u_{2}(X_{1},X_{2},X_{3})&\approx u_{2}^{(0)}(X_{1},X_{2})+X_{3}u_{2}^{(1)}(X_{1},X_{2})+\cdots,\\
u_{3}(X_{1},X_{2},X_{3})&\approx u_{3}^{(0)}(X_{1},X_{2})+X_{3}u_{3}^{(1)}(X_{1},X_{2})+\cdots,
\end{split}
\end{equation}
where $u_{1}(X_{1},X_{2},X_{3})$ and  $u_{2}(X_{1},X_{2},X_{3})$ are the in-plane displacements of the plate, and introduce the notation $\xi(X_{1} ,X_{2})=u_{3}^{(0)}(X_{1} ,X_{2} )$ to emphasise the fact that this out-of-plane component of the mid-surface $\mathsf{S}$ (where $X_{3}=0$) may be large compared to any other displacement components, then set $w(X_{1},X_{2},X_{3})=X_{3}u_{3}^{(1)}(X_{1},X_{2})+\cdots$.
The nematic plate in a deformed state is depicted in Figure~\ref{NLC:fig:Fig-plates}-bottom.  We further define a tensor $\textbf{g}=\textbf{G}-\textbf{I}$, with components $g_{ij}=G_{ij}-\delta_{ij}\ll 1$, $i,j=1,2,3$,where $\boldsymbol{\delta}=(\delta_{ij})_{i,j=1,2,3}$ is the Kronecker symbol. Since the plate is thin, we assume $\textbf{G}=\textbf{G}(X_{1},X_{2})$, and therefore, $\textbf{g}=\textbf{g}(X_{1},X_{2})$. As $\textbf{G}$ is symmetric, $\textbf{g}=\textbf{g}^{T}$.  

The length scales for our plate model based on assumptions (P1)-(P3) are as follows \cite{Dervaux:2009:DCBA,Xu:2020:XFY}:
\begin{itemize}
\item $\xi$ is of order $\gamma H$, where $1<\gamma\ll L/H$, i.e, out-of-plane displacements are small compared to the characteristic length $L$, but may be large relative to the thickness $H$;
\item $u_{1}$ and $u_{2}$ are of order $\xi^2/L$;
\item $g_{11}$, $g_{12}$, $g_{22}$ are of order $\xi^2/L^2$;
\item $g_{13}$ and $g_{23}$ are of order $\xi/L$;
\item $g_{33}=J-1-g_{11}-g_{22}+g_{13}^2+g_{23}^2$ is of order $\xi^2/L^2$.
\end{itemize}

In this case, the left Cauchy-Green tensor given by \eqref{NLC:eq:BC} takes the equivalent form
\begin{equation}\label{NLC:eq:BD}
\textbf{B}=\textbf{G}^{-1}\left(\textbf{D}\textbf{D}^{T}+\textbf{D}+\textbf{D}^{T}+\textbf{I}\right)\textbf{G}^{-1},
\end{equation}
and its in-plane components are approximated to leading orders as follows,
\begin{equation}\label{NLC:eq:B:app}
B^{(0)}_{\alpha\beta}=\frac{\partial u^{(0)}_{\alpha}}{\partial X_{\beta}}+\frac{\partial u^{(0)}_{\beta}}{\partial X_{\alpha}}-2g_{\alpha\beta}+g_{\alpha 3}g_{\beta 3}+\delta_{\alpha\beta}, \qquad \alpha,\beta=1,2.
\end{equation}
Then, the Almansi strain tensor defined by \eqref{NLC:eq:Almansi} is approximated by a tensor $\textbf{e}^{(0)}=\left(e^{(0)}_{ij}\right)_{i,j=1,2,3}$ with the in-plane components defined by
\begin{equation}\label{NLC:eq:e0}
e^{(0)}_{\alpha\beta}=\frac{1}{2}\left(\frac{\partial u^{(0)}_{\alpha}}{\partial X_{\beta}}+\frac{\partial u^{(0)}_{\beta}}{\partial X_{\alpha}}-2g_{\alpha\beta}+g_{\alpha 3}g_{\beta 3}\right),
\qquad \alpha,\beta=1,2.
\end{equation}
By assumption (P3), the Cauchy stress in the normal direction to the deformed mid-surface vanishes. As only moderate deflections are considered here (the plate is only slightly bent), the curvature of the plate is small and we can assume that the surface normal does not deviate much from $\textbf{e}_{3}$, so that the stress components $T_{13}$, $T_{23}$, $T_{33}$ are small compared to other stress components everywhere in the plate \cite{Dervaux:2008:DBA,Dervaux:2009:DCBA}. Then $T_{13}=T_{23}=0$ implies $e^{(0)}_{13}=e^{(0)}_{23}=0$, i.e.,
\begin{equation}
\frac{\partial u_{1}}{\partial X_{3}}\approx-\frac{\partial\xi}{\partial X_{1}}+2g_{13},\qquad
\frac{\partial u_{2}}{\partial X_{3}}\approx-\frac{\partial\xi}{\partial X_{2}}+2g_{23}.
\end{equation}
Hence,
\begin{equation}
u_{1}\approx u_{1}^{(0)}(X_{1},X_{2})-X_{3}\left(\frac{\partial\xi}{\partial X_{1}}-2g_{13}\right),\qquad
u_{2}\approx u_{2}^{(0)}(X_{1},X_{2})-X_{3}\left(\frac{\partial\xi}{\partial X_{2}}-2g_{23}\right),
\end{equation}
where $u_{1}^{(0)}$ and $u_{2}^{(0)}$ are of order $\xi^2/L$. In addition, $e^{(0)}_{33}\approx-e^{(0)}_{11}-e^{(0)}_{22}$, due to the incompressibility condition, $\det\textbf{B}=1$ (see Appendix~\ref{NLC:sec:append:bcs}). Since $T^{(nc)}_{33}=0$, the hydrostatic pressure satisfies
\begin{equation}\label{NLC:eq:p}
p=\mu B_{33},
\end{equation}
and, since $B_{33}$ does not depend on $X_{3}$, the coefficients in its approximation given by \eqref{NLC:eq:p0app} are
\begin{equation}\label{NLC:eq:p0}
p^{(0)}=\mu\left(2e^{(0)}_{33}-g_{13}^2-g_{23}^2+1\right),\quad  p^{(1)}=p^{(2)}=0.
\end{equation}
The following approximations are now obtained for the in-plane components of the Cauchy stress tensor defined by \eqref{NLC:eq:cauchy:iso}:
\begin{equation}\label{NLC:eq:T0}
\begin{split}
T^{(0)}_{11}&=2\mu\left(e^{(0)}_{11}-e^{(0)}_{33}+\frac{1}{2}g_{13}^2+\frac{1}{2}g_{23}^2\right)\approx 2\mu\left(2e^{(0)}_{11}+e^{(0)}_{22}+\frac{1}{2}g_{13}^2+\frac{1}{2}g_{23}^2\right),\\
T^{(0)}_{22}&=2\mu\left(e^{(0)}_{22}-e^{(0)}_{33}+\frac{1}{2}g_{13}^2+\frac{1}{2}g_{23}^2\right)\approx 2\mu\left(e^{(0)}_{11}+2e^{(0)}_{22}+\frac{1}{2}g_{13}^2+\frac{1}{2}g_{23}^2\right),\\
T^{(0)}_{12}&=T^{(0)}_{21}=2\mu e^{(0)}_{12}.
\end{split}
\end{equation}
This is a generalised form of Hooke's law.

Similarly, the right Cauchy-Green tensor given by \eqref{NLC:eq:BC} is equal to
\begin{equation}\label{NLC:eq:CD}
\textbf{C}=\left(\textbf{D}^{T}+\textbf{I}\right)\textbf{G}^{-2}\left(\textbf{D}+\textbf{I}\right),
\end{equation}
and its in-plane components are approximated to leading orders by
\begin{equation}\label{NLC:eq:C:app}
C^{(0)}_{\alpha\beta}=\frac{\partial u^{(0)}_{\alpha}}{\partial X_{\beta}}+\frac{\partial u^{(0)}_{\beta}}{\partial X_{\alpha}}+\frac{\partial \xi}{\partial X_{\alpha}}\frac{\partial \xi}{\partial X_{\beta}}-2g_{\alpha\beta}-g_{\alpha 3}g_{\beta 3}+\delta_{\alpha\beta}, \qquad \alpha,\beta=1,2.
\end{equation}
Thus, the Green-Lagrange strain tensor defined by \eqref{NLC:eq:GL} is approximated by $\textbf{E}^{(0)}=\left(E^{(0)}_{ij}\right)_{i,j=1,2,3}$ with the in-plane components (see also \cite[p.~51]{Landau:1986:LL})
\begin{equation}\label{NLC:eq:E0}
E^{(0)}_{\alpha\beta}=\frac{1}{2}\left(\frac{\partial u^{(0)}_{\alpha}}{\partial X_{\beta}}+\frac{\partial u^{(0)}_{\beta}}{\partial X_{\alpha}}+\frac{\partial \xi}{\partial X_{\alpha}}\frac{\partial \xi}{\partial X_{\beta}}-2g_{\alpha\beta}-g_{\alpha 3}g_{\beta 3}\right),
\qquad \alpha,\beta=1,2.
\end{equation}
By the incompressibility condition, $\det\textbf{C}=1$, it follows that $E^{(0)}_{33}\approx-E^{(0)}_{11}-E^{(0)}_{22}$ (see Appendix~\ref{NLC:sec:append:bcs}).

The in-plane components of the associated second Piola-Kirchhoff stress tensor given by \eqref {NLC:eq:2PK:iso} are then approximated as follows:
\begin{equation}\label{NLC:eq:S0}
\begin{split}
S^{(0)}_{11}&=2\mu\left(E^{(0)}_{11}-E^{(0)}_{33}+\frac{1}{2}g_{13}^2+\frac{1}{2}g_{23}^2\right)\approx 2\mu\left(2E^{(0)}_{11}+E^{(0)}_{22}+\frac{1}{2}g_{13}^2+\frac{1}{2}g_{23}^2\right),\\
S^{(0)}_{22}&=2\mu\left(E^{(0)}_{22}-E^{(0)}_{33}+\frac{1}{2}g_{13}^2+\frac{1}{2}g_{23}^2\right)\approx 2\mu\left(E^{(0)}_{11}+2E^{(0)}_{22}+\frac{1}{2}g_{13}^2+\frac{1}{2}g_{23}^2\right),\\
S^{(0)}_{12}&=S^{(0)}_{21}=2\mu E^{(0)}_{12}.
\end{split}
\end{equation}
This is the stress tensor that will be used to formulate the plate model. Next, we take into account the fact that the nematic director can re-orient.

\subsubsection{Director orientation}

When the deformation is not stress free, it is also necessary to determine the director orientation. By \eqref{NLC:eq:cauchy} and \eqref{NLC:eq:orient:matobj}, the following condition must be satisfied
\begin{equation}\label{NLC:eq:orient:plate}
\left(\textbf{g}\textbf{T}^{(0)}-\textbf{T}^{(0)}\textbf{g}\right)\textbf{n}=\textbf{0}.
\end{equation}
This is equivalent to
\begin{equation}\label{NLC:eq:orient:plate:T0}
\textbf{g}\textbf{T}^{(0)}=\textbf{T}^{(0)}\textbf{g},
\end{equation}
which, by \eqref{NLC:eq:T0}, it is also equivalent to
\begin{equation}\label{NLC:eq:orient:plate:e0}
\textbf{g}\textbf{e}^{(0)}=\textbf{e}^{(0)}\textbf{g}.
\end{equation}
Hence, 
\begin{equation}\label{NLC:eq:g13g23:zero}
g_{13}=g_{23}=0,
\end{equation}
i.e., the nematic director $\textbf{n}$ is tangent to the mid-surface, and
\begin{equation}\label{NLC:eq:orient:1}
g_{11}\varepsilon_{12}+g_{12}\varepsilon_{22}=g_{12}\varepsilon_{11}+g_{22}\varepsilon_{12},
\end{equation}
where
\begin{equation}\label{NLC:eq:vareps}
\varepsilon_{\alpha\beta}=\frac{1}{2}\left(\frac{\partial u^{(0)}_{\alpha}}{\partial X_{\beta}}+\frac{\partial u^{(0)}_{\beta}}{\partial X_{\alpha}}\right),
\qquad \alpha,\beta=1,2.
\end{equation}
The nematic director $\textbf{n}$ differs from the unit vector $\overline{\textbf{n}}$ obtained by the projection of the reference director $\textbf{n}_{0}$ on the mid-surface by a rotation $\textbf{R}$, i.e., $\textbf{n}=\textbf{R}\overline{\textbf{n}}$. Denoting by $\overline{\textbf{G}}=\textbf{I}+\overline{\textbf{g}}$ the `spontaneous' deformation tensor given by \eqref{NLC:eq:G:nu} with $\overline{\textbf{n}}$ instead of $\textbf{n}$, we obtain $\textbf{G}=\textbf{R}\overline{\textbf{G}}\textbf{R}^{-1}$ and $\textbf{g}=\textbf{R}\overline{\textbf{g}}\textbf{R}^{-1}$. Then, \eqref{NLC:eq:orient:plate:e0} takes the equivalent form
\begin{equation}\label{NLC:eq:orient:strain}
\textbf{R}\overline{\textbf{g}}\textbf{R}^{-1}\textbf{e}^{(0)}=\textbf{e}^{(0)}\textbf{R}\overline{\textbf{g}}\textbf{R}^{-1},
\end{equation}
from which the rotation $\textbf{R}$ can be found. To see this, we denote
\begin{equation}\label{NLC:eq:orient:gR}
\overline{\textbf{g}}=
\left[
\begin{array}{ccc}
\overline{g}_{1} & 0 & 0\\
0 & \overline{g}_{2} & 0\\
0 & 0 & \overline{g}_{3}
\end{array}
\right],\qquad
\textbf{R}=
\left[
\begin{array}{ccc}
\cos\alpha & \sin\alpha & 0\\
-\sin\alpha & \cos\alpha & 0\\
0 & 0 & 1
\end{array}
\right],
\end{equation}
so that
\begin{equation}\label{NLC:eq:orient:g}
g_{11}=\frac{\overline{g}_{1}+\overline{g}_{2}\tan^2\alpha}{1+\tan^2\alpha},\qquad
g_{22}=\frac{\overline{g}_{1}\tan^2\alpha+\overline{g}_{2}}{1+\tan^2\alpha},\qquad
g_{12}=\frac{\overline{g}_{2}\tan\alpha-\overline{g}_{1}\tan\alpha}{1+\tan^2\alpha}.
\end{equation}
By \eqref{NLC:eq:orient:1} and \eqref{NLC:eq:orient:g}, we obtain
\begin{equation}\label{NLC:eq:orient:3}
\varepsilon_{12}\tan^2\alpha-\left(\varepsilon_{11}-\varepsilon_{22}\right)\tan\alpha-\varepsilon_{12}=0,
\end{equation}
or equivalently,
\begin{equation}\label{NLC:eq:orient:alpha}
\tan(2\alpha)=\frac{2\varepsilon_{12}}{\varepsilon_{11}-\varepsilon_{22}},
\end{equation}
from which the angle $\alpha$ is found.

\subsubsection{The plate equations}

To obtain the plate equations, we compute the first variation of the energy given by \eqref{NLC:eq:deltaE}. In \eqref{NLC:eq:Eapp}, the stretching energy $E_{stretch}$, defined by \eqref{NLC:eq:Estretch}, involves both the in-plane displacements $u^{(0)}_{1}$, $u^{(0)}_{2}$ and the out-of-plane deflection $\xi$, and thus scales like $\mu H\xi^4/L^4$, while the bending energy $E_{bend}$, described by \eqref{NLC:eq:Ebend}, involves only the out-of-plane displacement $\xi$, and scales like $\mu H^3\xi^2/L^4$.
	
First, for the stretching energy given by \eqref{NLC:eq:Estretch}, we note (see Appendix B for details) that $E_{stretch}$ can be expressed in terms of stresses and strains as 
\begin{equation}\label{Estretch}
E_{stretch}= H\int_{-L_{2}/2}^{L_{2}/2}\int_{-L_{1}/2}^{L_{1}/2}S^{(0)}_{\alpha\beta}E^{(0)}_{\alpha\beta}dX_{1}dX_{2},
\end{equation}
where the Einstein summation notation for repeated indices is used. Then, the small variation with respect to the in-plane elastic strains takes the form (see also \cite[pp.~50-53]{Landau:1986:LL})
\begin{equation}\label{NLC:eq:varEstretch}
\begin{split}
\delta E_{stretch}&= H\int_{-L_{2}/2}^{L_{2}/2}\int_{-L_{1}/2}^{L_{1}/2}S^{(0)}_{\alpha\beta}\delta E^{(0)}_{\alpha\beta}dX_{1}dX_{2}\\
&= H\int_{-L_{2}/2}^{L_{2}/2}\int_{-L_{1}/2}^{L_{1}/2}S^{(0)}_{\alpha\beta}\left(\frac{\partial\delta u_{\alpha}^{(0)}}{\partial X_{\beta}}+\frac{\partial\xi}{\partial X_{\alpha}}\frac{\partial\delta\xi}{\partial X_{\beta}}\right)dX_{1}dX_{2}\\
&=-H\int_{-L_{2}/2}^{L_{2}/2}\int_{-L_{1}/2}^{L_{1}/2}\left[\frac{\partial S^{(0)}_{\alpha\beta}}{\partial X_{\beta}}\delta u_{\alpha}^{(0)}+\frac{\partial}{\partial X_{\beta}}\left(S^{(0)}_{\alpha\beta}\frac{\partial\xi}{\partial X_{\alpha}}\right)\delta\xi\right]dX_{1}dX_{2}.
\end{split}
\end{equation}
The last identity in \eqref{NLC:eq:varEstretch} was obtained through integration by parts. 

Second, we show in Appendix B that the bending energy $E_{bend}$ in \eqref{NLC:eq:Ebend} simplifies to 
\begin{equation}
E_{bend}=\frac{\mu H^3}{6}\int_{-L_{2}/2}^{L_{2}/2}\int_{-L_{1}/2}^{L_{1}/2} I_{1}^{(2)}dX_{1}dX_{2}.
\end{equation}
Then, the variation with respect to the out-of-plane deflection $\xi$ is equal to (see Appendix B for details and also \cite[pp.~38-44]{Landau:1986:LL})
\begin{equation}\label{NLC:eq:varEbend}
\begin{split}
\delta E_{bend}&=\frac{\mu H^3}{6}\int_{-L_{2}/2}^{L_{2}/2}\int_{-L_{1}/2}^{L_{1}/2}\delta\left(\Delta\xi\right)^2 dX_{1}dX_{2}+\mathcal{C}\\
&=\frac{\mu H^3}{3}\int_{-L_{2}/2}^{L_{2}/2}\int_{-L_{1}/2}^{L_{1}/2}\delta\xi\Delta^2\xi dX_{1}dX_{2}+\mathcal{C},
\end{split}
\end{equation}
where $\mathcal{C}$ represents contributions from the boundary conditions.

From \eqref{NLC:eq:Eapp}, \eqref{NLC:eq:varEstretch} and \eqref{NLC:eq:varEbend}, by the principle of stationary potential energy \cite[p.~306]{Ogden:1997}, assuming that external stretching forces can be neglected, the following equations are obtained for the equilibrium of the nematic plate under the external bending force $f_{P}$ acting in the direction normal to the plate, 
\begin{eqnarray}
&&\frac{\mu}{3}H^3\Delta^2\xi-H\frac{\partial}{\partial X_{\beta}}\left(S^{(0)}_{\alpha\beta}\frac{\partial\xi}{\partial X_{\alpha}}\right)=f_{P},\label{NLC:plate:1of3}\\
&&\frac{\partial S^{(0)}_{\alpha\beta}}{\partial X_{\beta}}=0,\qquad \alpha=1,2. \label{NLC:plate:23of3}
\end{eqnarray}
In these equations, $\mu H^{3}/3$ represents the bending modulus of the plate, and the unknowns are the out-of-plane bending $\xi$ and the in-plane stretching components $u^{(0)}_{1}$, $u^{(0)}_{2}$.

The static equilibrium of the nematic plate, describing both its Gaussian and the mean curvature, is fully characterised by equations \eqref{NLC:plate:1of3}-\eqref{NLC:plate:23of3} together with \eqref{NLC:eq:orient:g} and \eqref{NLC:eq:orient:alpha}, and completed by boundary conditions imposed on the `edge' of the plate.

A classic way to simplify these equations is to define the Airy stress function $\chi$, such that
\begin{equation}\label{NLC:eq:Airy}
S^{(0)}_{11}=\frac{\partial^2\chi}{\partial X_{2}^2},\qquad
S^{(0)}_{12}=-\frac{\partial^2\chi}{\partial X_{1}\partial X_{2}},\qquad
S^{(0)}_{22}=\frac{\partial^2\chi}{\partial X_{1}^2}.
\end{equation}
Denoting
\begin{equation}\label{NLC:eq:chixi}
[\chi,\xi]=\frac{1}{2}\frac{\partial^2\chi}{\partial X_{1}^2}\frac{\partial^2\xi}{\partial X_{2}^2}
+\frac{1}{2}\frac{\partial^2\chi}{\partial X_{2}^2}\frac{\partial^2\xi}{\partial X_{1}^2}
-\frac{\partial^2\chi}{\partial X_{1}\partial X_{2}}\frac{\partial^2\xi}{\partial X_{1}\partial X_{2}},
\end{equation}
and noting that the Gaussian curvature is approximately equal to
\begin{equation}\label{NLC:eq:gcurvature}
[\xi,\xi]=\frac{\partial^2\xi}{\partial X_{1}^2}\frac{\partial^2\xi}{\partial X_{2}^2}-\left(\frac{\partial^2\xi}{\partial X_{1}\partial X_{2}}\right)^2,
\end{equation}
we obtain
\begin{equation}\label{NLC:eq:Airy:2harmon:PsiG}
\begin{split}
\Delta^2\chi
&=\frac{\partial^4\chi}{\partial X_{1}^4}+2\frac{\partial^4\chi}{\partial X_{1}^2\partial X_{2}^2}+\frac{\partial^4\chi}{\partial X_{2}^4}\\
&=\left(\frac{\partial^4\chi}{\partial X_{2}^4}-\frac{1}{2}\frac{\partial^4\chi}{\partial X_{1}^2\partial X_{2}^2}\right)
+\left(\frac{\partial^4\chi}{\partial X_{1}^4}-\frac{1}{2}\frac{\partial^4\chi}{\partial X_{1}^2\partial X_{2}^2}\right)
+3\frac{\partial^4\chi}{\partial X_{1}^2\partial X_{2}^2}\\
&=\frac{\partial^2}{\partial X_{2}^2}\left(S^{(0)}_{11}-\frac{1}{2}S^{(0)}_{22}\right)
+\frac{\partial^2}{\partial X_{1}^2}\left(S^{(0)}_{22}-\frac{1}{2}S^{(0)}_{11}\right)
-3\frac{\partial^2S^{(0)}_{12}}{\partial X_{1}\partial X_{2}}\\
&=3\mu\left(\frac{\partial^2E^{(0)}_{11}}{\partial X_{2}^2}\
+\frac{\partial^2E^{(0)}_{22}}{\partial X_{1}^2}
-2\frac{\partial^2E^{(0)}_{12}}{\partial X_{1}\partial X_{2}}\right)\\
&=-3\mu\left([\xi,\xi]-\Psi_{G}\right),
\end{split}
\end{equation}
where
\begin{equation}\label{NLC:eq:PsiG}
\Psi_{G}=2\frac{\partial^2g_{12}}{\partial X_{1}\partial X_{2}}-\frac{\partial^2g_{11}}{\partial X_{2}^2}
-\frac{\partial^2g_{22}}{\partial X_{1}^2}
\end{equation}
is the source of Gaussian curvature, with $g_{11}$, $g_{22}$, $g_{12}$ described by \eqref{NLC:eq:orient:g}, and the angle $\alpha$ given by equation \eqref{NLC:eq:orient:alpha}.

Using \eqref{NLC:eq:Airy}, \eqref{NLC:eq:chixi}, and \eqref{NLC:eq:Airy:2harmon:PsiG}, the plate equations now take the equivalent form
\begin{eqnarray}
&&\frac{\mu}{3}H^3\Delta^2\xi-2H[\chi,\xi]=f_{P},\label{NLC:plate:1of2}\\
&&\Delta^2\chi+3\mu\left([\xi,\xi]-\Psi_{G}\right)=0,\label{NLC:plate:2of2}
\end{eqnarray}
where $\Psi_{G}$ is given by \eqref{NLC:eq:PsiG}.

These equations are valid for nematic liquid crystalline plates with `free' nematic director, subjected simultaneously to applied forces and optothermal stimuli. A direct application of these equations would be the wrinkling or pattern formation in nematic thin films \cite{Agrawal:2012:etal,Huang:2005:HHS,Krieger:2019:KD,Plucinsky:2017:PB,Soni:2016:SPP}.

\subsection{Frozen nematic director}\label{NLC:sec:nfrozen}

When the nematic director is `frozen', the plate equations are obtained in a similar manner using the modified strain-energy function given by \eqref{NLC:eq:Wnc:NH:q}. The difference is that, for this case, if the deformation is not stress free, then a condition analogous to \eqref{NLC:eq:orient:plate} must be derived from 
\eqref{NLC:eq:cauchy:frozen} and \eqref{NLC:eq:orient:matobj:frozen}.

\section{Applications: Nematic rings}\label{NLC:sec:rings}

For illustration, we apply the plate equations to some classical problems of temperature-driven shape changes based on the disc geometry. We focus our attention on nematic circular annular discs, or rings, with the `frozen' director uniformly distributed throughout the thickness and circularly symmetric around the centre. These rings can deform homogeneously through the thickness and inhomogeneously in the plane \cite{Agrawal:2012:etal,Ahn:2015:ALC,Ahn:2016:etal,deHaan:2014:etal,Efrati:2009:ESK,Konya:2016:KGPS,Kowalski:2017:etal,Modes:2010:MBW,Modes:2011:MBW,Modes:2011:MW,Mostajeran:2015,Mostajeran:2016:MWWW,Pismen:2014,Warner:2018:WM}. For the thin plate model, in the absence of external forces, when only heat generated or optomechanical responses arise, the `spontaneous' deformation tensor is equal to $\textbf{G}=\text{diag}\left(1+g_{1}, 1+g_{2}, 1\right)$ in cylindrical polar coordinates $(R,\Theta,Z)$, and the components of the stress tensor defined by \eqref{NLC:eq:S0} are equal to zero, i.e., the deformation is stress free \cite{Efrati:2009:ESK,Efrati:2013:ESK}. If $g_{1}=g_{2}$, then $a=1$ and the ring remains flat.

\subsection{Large out-of-plane deformations}\label{NLC:sec:ring1}

When stress-free out-of-plane deformations are much larger than the thickness of the plate, i.e., $\xi\gg H$, bending can be neglected, and the only equation that remains to be solved is the Monge-Amp\`{e}re equation \cite{Dervaux:2008:DBA,Xu:2020:XFY}
\begin{equation}\label{NLC:plate:sol2}
[\xi,\xi]=\Psi_{G},
\end{equation}
where, in polar coordinates,
\begin{equation}
[\xi,\xi]=\frac{1}{R}\frac{\partial\xi}{\partial R}\frac{\partial^2\xi}{\partial R^2}+\frac{1}{R^2}\frac{\partial^2\xi}{\partial R^2}\frac{\partial^2\xi}{\partial\Theta^2}-\frac{1}{R^2}\left(\frac{\partial^2\xi}{\partial R\partial\Theta}-\frac{1}{R}\frac{\partial\xi}{\partial\Theta}\right)^2
\end{equation}
and 
\begin{equation}
\Psi_{G}=(g_{1}-g_{2})\frac{\delta(R)}{R},
\end{equation}
with $\delta(\cdot)$ denoting the Dirac delta function. 

\begin{figure}[ht]
	\begin{center}
		\includegraphics[width=0.65\textwidth]{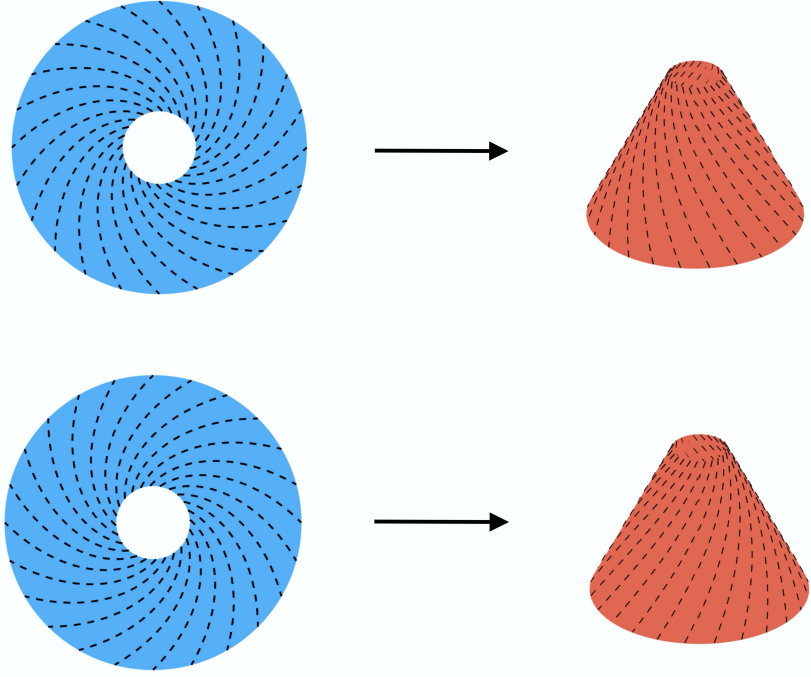}
		\caption{Schematics of natural shapes taken by a nematic ring with $\textbf{n}_{0}=[\cos\varphi, \sin\varphi, 0 ]^{T}$ in cylindrical polar coordinates $(R,\Theta,Z)$, where $\varphi=\pi/4$ (top) or $\varphi=-\pi/4$ (bottom), as truncated right circular cones when heated from an initial low temperature. The undeformed bodies at low temperature are represented in blue colour, and their deformed shapes at high temperature are shown in red colour.}\label{NLC:fig:spiral-cone}
	\end{center}
\end{figure}

We consider rings with the nematic field forming spirals around the centre, which are initially at low temperature with $0<a<1$, then heated \cite{Kassianidis:2008:KOMP,Klein:2007:KES,Modes:2011:MBW}. We set $\textbf{n}_{0}=[\cos\varphi, \sin\varphi, 0 ]^{T}$, where $-\pi/2<\varphi<\pi/2$, in cylindrical polar coordinates $(R,\Theta,Z)$, and look for solutions of the form $\xi(R,\Theta)=\xi(R)$, which leads to
\begin{equation}\label{NLC:eq:sol:cone3}
\xi(R,\Theta)=(R-R_{0})a^{1/3}\sqrt{\frac{a^{2(\nu+1)/3}+6+a^{-2(\nu+1)/3}}{2\left(a^{2(\nu+1)/3}+1\right)}}.
\end{equation}
The resulting shape is a truncated right circular cone (see Figure~\ref{NLC:fig:spiral-cone}).

\subsection{Moderate out-of-plane deformations}\label{NLC:sec:ring2}

Next, we assume that the nematic director is aligned in the azimuthal direction, i.e., $\textbf{n}_{0}=[0, 1, 0 ]^{T}$ in cylindrical polar coordinates $(R,\Theta,Z)$, and the ring is initially at high temperature with $a>1$, then cooled, so that the material will tend to elongate in the azimuthal direction and contract in the radial direction. In this case, when the out-of-plane deformations are of the same order as the plate thickness, i.e., $\xi\approx H$, a stability analysis similar to that carried out in \cite{Dervaux:2008:DBA} can be performed.

For the nematic ring, at the inner surface, where $R=A$, we assume $u_{1}=0$, $u_{2}=0$ and $\xi=0$, such that $\partial\xi/\partial R=0$, while at the outer surface, where $R=B$, we impose $S^{(0)}_{11}=2\mu s_{0}$, with $s_{0}$ constant, and $S^{(0)}_{12}=0$, and also the two boundary conditions that the ‘edge’ of the plate is free from traction (see Appendix~\ref{NLC:sec:append:bcs}).

We are interested in deformations with the in-plane displacement and stresses independent of $\Theta$, i.e., $u_{1}=u_{1}(R)$, $u_{2}=u_{2}(R)$, and $S^{(0)}_{11}=S^{(0)}_{11}(R)$, $S^{(0)}_{12}=S^{(0)}_{12}(R)$, $S^{(0)}_{22}=S^{(0)}_{22}(R)$, while the out-of-plane displacement may vary both with $R$ and $\Theta$, i.e., $\xi=\xi(R,\Theta)$.

Under these conditions, after rescaling, so that $R\to R/B$, and denoting the ratio between inner and outer radii by $R_{0}=A/B$, so that $R\in(R_{0},1)$, we can write
\begin{equation}\label{NLC:eq:u:app}
u_{1}(R)=u_{1}(R_{0})+u_{1}'(R_{0})(R-R_{0}), \qquad u_{2}(R)=u_{2}(R_{0})+u_{2}'(R_{0})(R-R_{0}),
\end{equation}
where $u_{1}'$ and $u_{2}'$ represent the first derivative of $u_{1}$ and $u_{2}$ with respect to $R$, respectively. We then solve equations \eqref{NLC:plate:23of3}, given the constitutive relations \eqref{NLC:eq:S0}, and obtain (see Appendix~\ref{NLC:sec:append:ring} for details) the stress components
\begin{equation}\label{NLC:eq:S0:sol}
S^{(0)}_{11}=2\mu s_{0}, \qquad S^{(0)}_{22}=-2\mu s_{0},\qquad S^{(0)}_{12}=0,
\end{equation}	
and the displacement components
\begin{equation}\label{NLC:eq:u:sol}
u_{1}=\left(R-R_{0}\right)\left(g_{1}+g_{2}\right),\qquad u_{2}=0,
\end{equation}
for all $R\in(R_{0},1)$. Such situation may arise, for example, when the ring is continuously attached to an inner disc that remains undeformed \cite{Ahn:2015:ALC}.

\begin{figure}[ht]
	\begin{center}
		\includegraphics[width=0.75\textwidth]{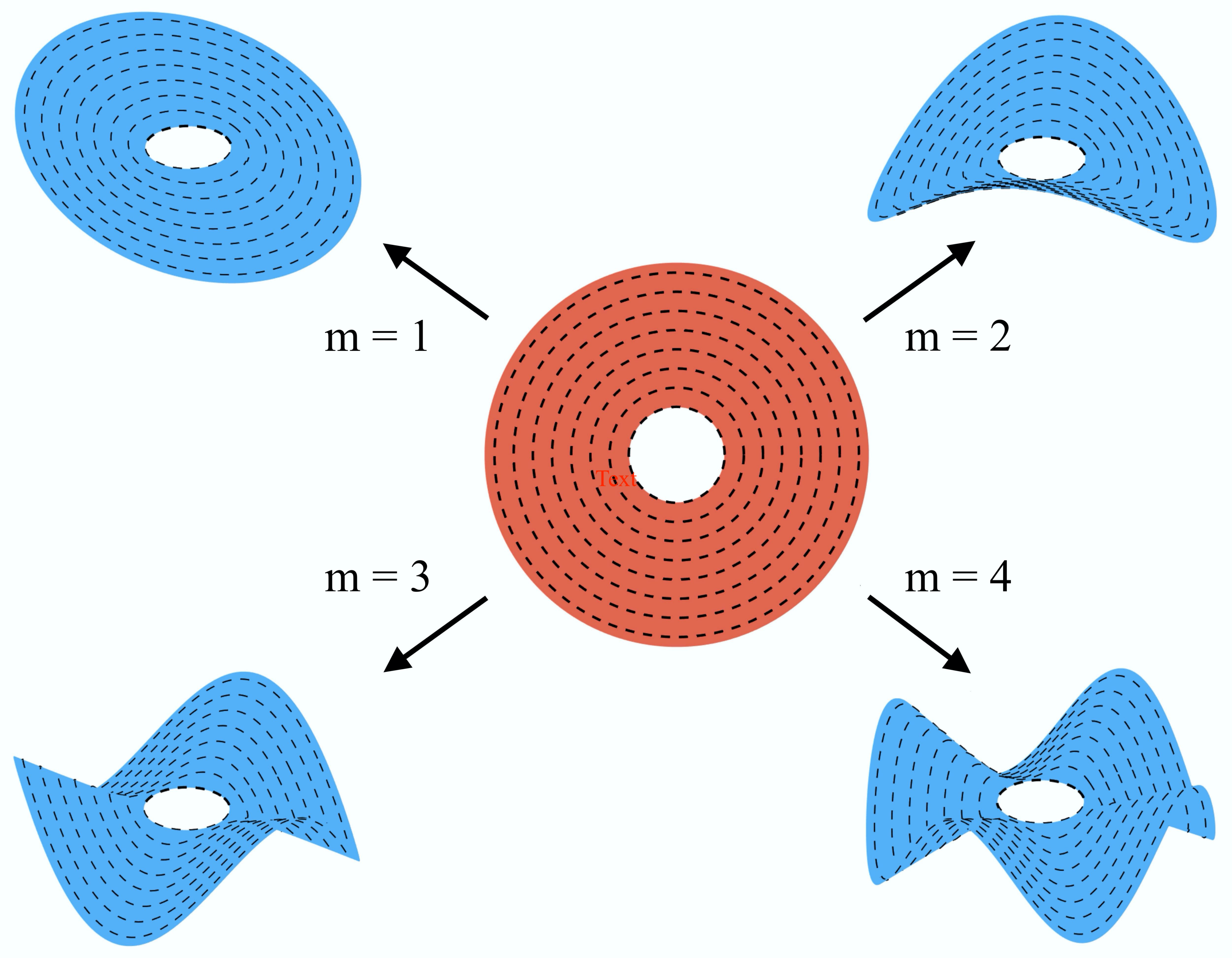}
		\caption{Schematics of natural shapes taken by a nematic ring with $\textbf{n}_{0}=[0, 1, 0 ]^{T}$ in cylindrical polar coordinates $(R,\Theta,Z)$ as d-cones (figures on top-left, top-right, bottom-left, bottom-right showing modes $m=1, 2,3,4$, respectively) when cooled from an initial high temperature (middle figure). The undeformed body at high temperature is represented in red colour, and the deformed shapes at low temperature are shown in blue colour.}\label{NLC:fig:dcones}
	\end{center}
\end{figure}

To find the out-of-plane deformations, we solve equation \eqref{NLC:plate:1of3}, which is equivalent to
\begin{equation}\label{NLC:eq:plate}
\Delta^2\xi-\frac{3}{H^2\mu}\left(S^{(0)}_{11}\frac{\partial^2\xi}{\partial R^2}+S^{(0)}_{22}\frac{\partial^2\xi}{R^2\partial\Theta^2}\right)=0,
\end{equation}
where the biharmonic operator of the out-of-plane displacement in polar coordinates takes the form:
\begin{equation}
\Delta^2\xi=\frac{\partial^4\xi}{\partial R^4}+\frac{2\partial^4\xi}{R^2\partial R^2\partial\Theta^2}+\frac{\partial^4\xi}{R^4\partial\Theta^4}+\frac{2\partial^3\xi}{R\partial R^3}-\frac{2\partial^3\xi}{R^3\partial R\partial^2\Theta}-\frac{\partial^2\xi}{R^2\partial R^2}+\frac{4\partial^2\xi}{R^4\partial\Theta^2}+\frac{\partial\xi}{R^3\partial R}.
\end{equation}

The four boundary conditions for equation \eqref{NLC:eq:plate} are as follows:
\begin{itemize}
\item At the inner surface, where $R=R_{0}$,
\begin{equation}\label{NLC:eq:bc12}
\xi=0,\qquad \frac{\partial\xi}{\partial R}=0.
\end{equation}
\item At the outer surface, where $R=1$, setting $\widetilde{\textbf{n}}=[1,0]^{T}$ and $\widetilde{\textbf{t}}=[0,1]^{T}$ as the outward unit normal and tangent vector, respectively, the two equations \eqref{NLC:eq:bc3:app}-\eqref{NLC:eq:bc4:app} take the form
\begin{equation}\label{NLC:eq:bc34}
\frac{\partial^2\xi}{\partial R^2}=0,\qquad
\frac{\partial^3\xi}{\partial R^3}=0.
\end{equation}
\end{itemize}

Changing variable to $\rho=R-R_{0}$ and assuming solutions of the form $\xi=\zeta(\rho)\cos(m\Theta)$, equation \eqref{NLC:eq:plate} reduces to the following fourth-order ordinary differential equation,
\begin{equation}\label{NLC:eq:4ode}
\rho^4\zeta^{iv}+2\rho^3\zeta'''-\left(1+2m^2+C\rho^2\right)\rho^2\zeta''+\left(1+2m^2\right)\rho\zeta'-\left(4-m^2+C\rho^2\right)m^2\zeta=0,
\end{equation}
where $\zeta'$, $\zeta''$, $\zeta'''$, and $\zeta^{iv}$ are the derivatives of order 1, 2, 3 and 4, respectively, of $\zeta$ with respect to $\rho$, and $C=3S^{(0)}_{11}/(H^2\mu)=-3S^{(0)}_{22}/(H^2\mu)=6s_{0}/H^2>0$ is a control parameter.

For equation \eqref{NLC:eq:4ode}, the boundary conditions are:
\begin{itemize}
	\item At $\rho=0$, 
	\begin{equation}\label{NLC:eq:bc12:ode}
	\zeta=0,\qquad\zeta'=0.
	\end{equation}
	\item At $\rho=1-R_{0}$, 
	\begin{equation}\label{NLC:eq:bc34:ode}
	\zeta''=0,\qquad
	\zeta'''=0.
	\end{equation}
\end{itemize}

\begin{figure}[htbp]
	\begin{center}
		\subfigure[]{\includegraphics[width=0.50\textwidth]{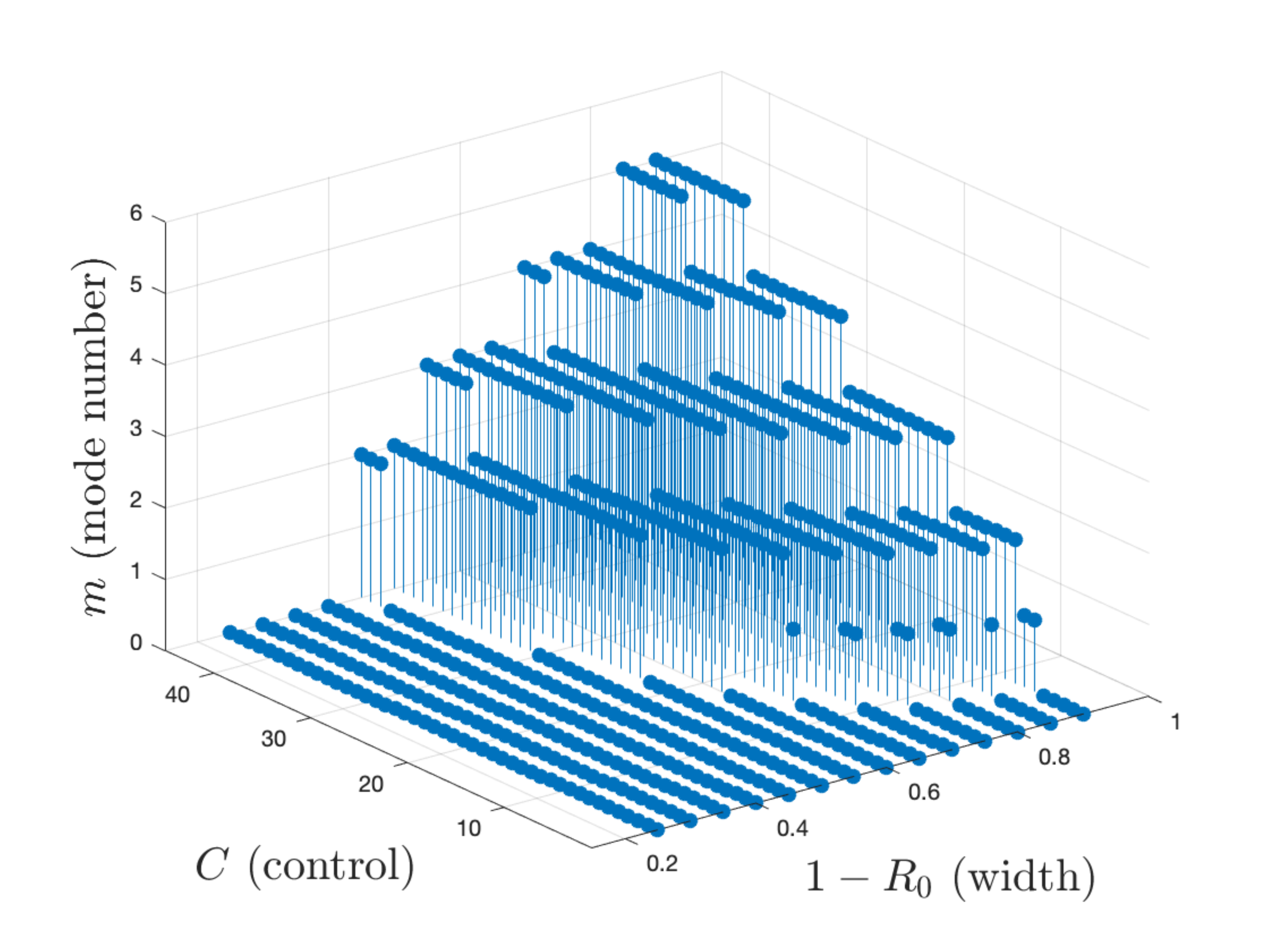}}
		\subfigure[]{\includegraphics[width=0.45\textwidth]{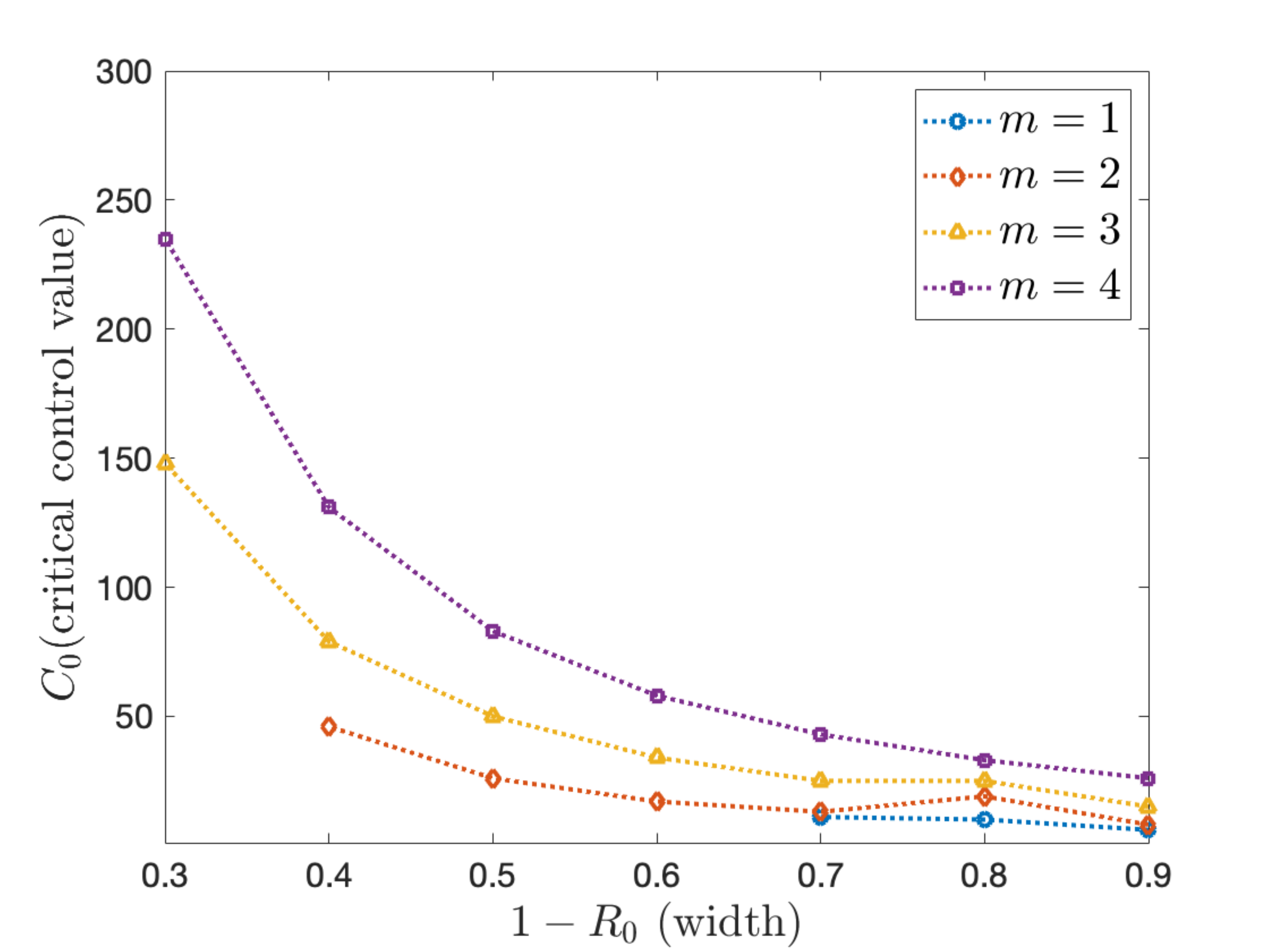}}
		\caption{(a) The effect of variation in the control parameter $C=6s_{0}/H^2>0$ on the mode number $m$ for rings with different ratios $R_{0}$ between the inner and outer radii; (b) Critical values $C_{0}$ of the control parameter at which mode numbers $m\in\{1,2,3,4\}$ are observed as the width of the ring $1-R_{0}$ changes.}\label{NLC:fig:mode}
	\end{center}
\end{figure}

For $m=0$, the ring deforms into a truncated right circular cone. For $m>0$, by treating the boundary value problem as an eigenvalue problem in the control parameter $C$, we can find the threshold value for this parameter such that a non-trivial out-of-plane deformation occurs. In this case, the evolved shapes are anti-cones (developable or d-cones) as shown in Figure~\ref{NLC:fig:dcones}.

To solve equation \eqref{NLC:eq:4ode}, subject to the boundary conditions \eqref{NLC:eq:bc12:ode}-\eqref{NLC:eq:bc34:ode}, we apply a numerical scheme described in \cite[pp.~460-461]{goriely17}. Figure~\ref{NLC:fig:mode}(a) shows the mode number $m$ observed for rings with different ratios between the inner and outer radii, $R_{0}$, as the prescribed (control) value $C=6s_{0}/H^2>0$ changes. In Figure~\ref{NLC:fig:mode}(b), for mode numbers $m=1,2,3,4$, the minimum critical control values $C_{0}$ are plotted as functions of the width, $1-R_{0}$. Note that, for width $1-R_{0}\in\{0.4, 0.5, 0.6\}$, the mode number $m=1$ is not obtained, while for width $1-R_{0}=0.3$, mode numbers $m=1$ and $m=2$ are not found. 

\section{Conclusion}\label{NLC:sec:conclude}

We have developed here a F\"{o}ppl-von K\'{a}rm\'{a}n-type model to describe the elastic behaviour of NLC plates, subject to combined optothermal stimulation and mechanical loading. We  achieve this by exploiting the multiplicative decomposition of the deformation gradient into an elastic component and a `spontaneous' deformation tensor. To illustrate the application of this model, we provide analytical solutions to combined natural and forced shape changes of circular rings with `frozen' nematic director in circular disclination patterns, where the deformation is homogeneous through the thickness and the director remains tangent to the mid-surface. 

Stress free natural shape changes had been analysed extensively from the kinematic point of view before \cite{deHaan:2014:etal,Modes:2016:MW,Warner:2020}. However, the ability of these complex morphing materials to do physical work (which is critical in soft robotics applications, for instance) has been much less addressed \cite{deHaan:2014:etal,Tajbakhsh:2001:TT}. As interest in synthetic morphogenic materials continues to increase \cite{Cicconofri:2020:etal,Noselli:2019:etal}, the proposed model is timely and accounts for both the displacement and stress fields in a deforming nematic plate. 

Our derivation is based on the neoclassical strain-energy function describing an ideal liquid crystal solid with `free' or `frozen' nematic director, but the dimensional reduction procedure could, in principle, be extended to other strain-energy functions as well \cite{DeSimone:2009:dST,Fried:2004:FS,Mihai:2020a:MG}. The equilibrium plate equations derived here could further be employed to tackle inverse problems where, for a target deformation, the orientation of the nematic director and the forces to be applied in the reference configuration are sought \cite{Aharoni:2014:ASK,Griniasty:2019:GAE,Mostajeran:2016:MWWW,Plucinsky:2016:PLB,Siefert:2019:etal,Warner:2018:WM}.

\paragraph{Acknowledgement.} We thank Fehmi Cirak (University of Cambridge) for a discussion on constitutive modelling of liquid crystalline solids. The support by the Engineering and Physical Sciences Research Council of Great Britain under research grants EP/R020205/1 to Alain Goriely and EP/S028870/1 to L. Angela Mihai is gratefully acknowledged.

\appendix
\setcounter{equation}{0}
\renewcommand{\theequation}{A.\arabic{equation}}

\section{Relations between stress tensors}\label{NLC:sec:append:stress}

In this appendix, we derive the stress tensors of the deformed nematic material in terms of the stresses in the base polymeric network when the nematic director is `free' to rotate relative to the elastic matrix, and when the nematic director is `frozen'.

\subsection{Free director}

When the nematic director is `free', $\textbf{F}$ and $\textbf{n}$ are independent variables, and the strain-energy function given by \eqref{NLC:eq:Wnc:NH} takes the equivalent form
\begin{equation}\label{NLC:eq:Wnc:NH:free}
\mathsf{W}^{(nc)}(\lambda_{1},\lambda_{2},\lambda_{3})=\frac{\mu}{2}\left\{a^{2\nu/3}\left[\lambda_{1}^2+\lambda_{2}^2+\lambda_{3}^2-\left(1- a^{-2(1+\nu)/3}\right)\textbf{n}\cdot\left(\sum_{i=1}^3\lambda_{i}^2\textbf{e}_{i}\otimes\textbf{e}_{i}\right)\textbf{n}\right]-3\right\},
\end{equation}
where $\{\lambda_{i}^2\}_{i=1,2,3}$ and $\{\textbf{e}_{i}\}_{i=1,2,3}$ denote the eigenvalues and eigenvectors, respectively, of the left Cauchy-Green tensor, such that
\begin{equation}\label{NLC:eq:FFt}
\textbf{F}\textbf{F}^{T}=\sum_{i=1}^3\lambda_{i}^2\textbf{e}_{i}\otimes\textbf{e}_{i}.
\end{equation}

Next, we use the multiplicative decomposition \eqref{NLC:eq:FGA} of $\textbf{F}$ to obtain the stress tensors of the nematic material in terms of the stresses of the elastically deformed polymeric network. Since $\textbf{G}$ is symmetric, the Cauchy stress tensor for the nematic material with the strain-energy function described by \eqref{NLC:eq:Wnc:NH} is calculated as follows,
\begin{equation}\label{NLC:eq:cauchy}
\begin{split}
\textbf{T}^{(nc)}
&=J^{-1}\frac{\partial W^{(nc)}}{\partial\textbf{F}}\textbf{F}^{T}-p^{(nc)}\textbf{I}\\
&=J^{-1}\textbf{G}^{-T}\frac{\partial W}{\partial\textbf{A}}\textbf{A}^{T}\textbf{G}^{T}-p^{(nc)}\textbf{I}\\
&=J^{-1}\textbf{G}^{-1}\frac{\partial W}{\partial\textbf{A}}\textbf{A}^{T}\textbf{G}-p^{(nc)}\textbf{I}\\
&=J^{-1}\textbf{G}^{-1}\textbf{T}\textbf{G},
\end{split}
\end{equation}
where $\textbf{T}$ is the elastic Cauchy stress defined by \eqref{NLC:eq:cauchy:iso}, $J=\det\textbf{F}=\det\textbf{G}\det\textbf{A}=\det\textbf{G}=a^{(1-2\nu)/3}$, and the scalar $p^{(nc)}$ (the hydrostatic pressure) represents the Lagrange multiplier for the internal constraint $J=a^{(1-2\nu)/3}$.

Note that the Cauchy stress tensor $\textbf{T}^{(nc)}$ given by \eqref{NLC:eq:cauchy} is not symmetric in general \cite{Anderson:1999:ACF,Soni:2016:SPP,Zhang:2019:etal}.  In addition, the following condition must hold \cite{Anderson:1999:ACF,Zhang:2019:etal},
\begin{equation}\label{NLC:eq:orient}
\frac{\partial W^{(nc)}}{\partial\textbf{n}}=\textbf{0},
\end{equation}
or equivalently, by the principle of material objectivity stating that constitutive equations must be invariant under changes of frame of reference (see \cite{Anderson:1999:ACF} for details),
\begin{equation}\label{NLC:eq:orient:matobj}
\left(\textbf{T}^{(nc)T}-\textbf{T}^{(nc)}\right)\textbf{n}=\textbf{0}.
\end{equation}

The first Piola-Kirchhoff stress tensor for the nematic material is equal to
\begin{equation}\label{NLC:eq:1PK:nc}
\textbf{P}^{(nc)}=\textbf{T}^{(nc)}\text{Cof}(\textbf{F})=\textbf{G}^{-1}\textbf{T}\textbf{A}^{-T}=\textbf{G}^{-1}\textbf{P},
\end{equation}
where $\textbf{P}$ is the elastic first Piola-Kirchhoff stress given by \eqref{NLC:eq:1PK:iso}.

The corresponding second Piola-Kirchhoff stress tensor is
\begin{equation}\label{NLC:eq:2PK:nc}
\textbf{S}^{(nc)}=\textbf{F}^{-1}\textbf{P}^{(nc)}=\textbf{A}^{-1}\textbf{G}^{-2}\textbf{P}=\textbf{A}^{-1}\textbf{G}^{-2}\textbf{A}\textbf{S},
\end{equation}
where $\textbf{S}$ is the elastic second Piola-Kirchhoff stress tensor given by \eqref{NLC:eq:2PK:iso}.

\subsection{Frozen director}

If the nematic director is `frozen', then the strain-energy function described by \eqref{NLC:eq:Wnc:NH} can be written equivalently as follows,
\begin{equation}\label{NLC:eq:Wnc:NH:frozen}
\mathsf{W}^{(nc)}(\lambda_{1},\lambda_{2},\lambda_{3})=\frac{\mu}{2}\left\{a^{2\nu/3}\left[\lambda_{1}^2+\lambda_{2}^2+\lambda_{3}^2-\left(1-a^{-2(1+\nu)/3}\right)\frac{\left|\left(\sum_{i=1}^3\lambda_{i}^2\textbf{E}_{i}\otimes\textbf{E}_{i}\right)\textbf{n}_{0}\right|^2}{\textbf{n}_{0}\cdot\left(\sum_{i=1}^3\lambda_{i}^2\textbf{E}_{i}\otimes\textbf{E}_{i}\right)\textbf{n}_{0}}\right]-3\right\},
\end{equation}
where $\{\lambda_{i}^2\}_{i=1,2,3}$ and $\{\textbf{E}_{i}\}_{i=1,2,3}$ represent the eigenvalues and eigenvectors, respectively, of the right Cauchy-Green tensor, such that
\begin{equation}\label{NLC:eq:FtF}
\textbf{F}^{T}\textbf{F}=\sum_{i=1}^3\lambda_{i}^2\textbf{E}_{i}\otimes\textbf{E}_{i}.
\end{equation}
To obtain the above form, we have used the identities
\begin{equation}\label{NLC:eq:I4:frozen}
\textbf{n}\cdot\textbf{F}\textbf{F}^{T}\textbf{n}
=\frac{|\textbf{F}^{T}\textbf{F}\textbf{n}_{0}|^2}{|\textbf{F}\textbf{n}_{0}|^2}
=\frac{|\textbf{F}^{T}\textbf{F}\textbf{n}_{0}|^2}{\textbf{n}_{0}\cdot\textbf{F}^{T}\textbf{F}\textbf{n}_{0}}.
\end{equation}
The kinematic interpretation of $\textbf{n}_{0}\cdot\textbf{F}^{T}\textbf{F}\textbf{n}_{0}=|\textbf{F}\textbf{n}_{0}|^2$ is that it represents the square of the stretch ratio in the direction $\textbf{n}_{0}$.

In this case also, we can express the stress tensors of the nematic material in terms of the stresses in the deformed polymeric network. First, we define a modified strain-energy function with independent variables $\textbf{F}$ and $\textbf{n}$,
\begin{equation}\label{NLC:eq:Wnc:NH:q}
\widehat{W}^{(nc)}(\textbf{F},\textbf{n})
=\frac{\mu}{2}\left\{a^{2\nu/3}\left[\text{tr}\left(\textbf{F}\textbf{F}^{T}\right)-\left(1- a^{-2(1+\nu)/3}\right)\textbf{n}\cdot\textbf{F}\textbf{F}^{T}\textbf{n}\right]-3\right\}-q\left(\textbf{n}\cdot\frac{\textbf{F}\textbf{n}_{0}}{|\textbf{F}\textbf{n}_{0}|}-1\right),
\end{equation}
where the scalar $q$ is the Lagrange multiplier for the constraint \eqref{NLC:eq:n0n}.

Then, the Cauchy stress tensor for the nematic material takes the form
\begin{equation}\label{NLC:eq:cauchy:frozen}
\begin{split}
\widehat{\textbf{T}}^{(nc)}
&=J^{-1}\frac{\partial\widehat{W}^{(nc)}}{\partial\textbf{F}}\textbf{F}^{T}-p^{(nc)}\textbf{I}\\
&=J^{-1}\textbf{G}^{-1}\textbf{T}\textbf{G}-J^{-1}q\left(\textbf{I}-\frac{\textbf{F}\textbf{n}_{0}\otimes\textbf{F}\textbf{n}_{0}}{|\textbf{F}\textbf{n}_{0}|^2}\right)\textbf{n}\otimes\frac{\textbf{F}\textbf{n}_{0}}{|\textbf{F}\textbf{n}_{0}|},
\end{split}
\end{equation}
where $\textbf{T}$ is the elastic Cauchy stress defined by \eqref{NLC:eq:cauchy:iso}, and $p^{(nc)}$ is the Lagrange multiplier for the volume constraint $J=a^{(1-2\nu)/3}$.

As the Cauchy stress tensor given by \eqref{NLC:eq:cauchy:frozen} is not symmetric in general, the following additional condition must hold,
\begin{equation}\label{NLC:eq:orient:frozen}
\frac{\partial\widehat{W}^{(nc)}}{\partial\textbf{n}}=\textbf{0},
\end{equation}
or equivalently,
\begin{equation}\label{NLC:eq:orient:matobj:frozen}
\left(\widehat{\textbf{T}}^{(nc)T}-\widehat{\textbf{T}}^{(nc)}\right)\textbf{n}=\textbf{0}.
\end{equation}
The corresponding first Piola-Kirchhoff stress tensor for the nematic material is equal to
\begin{equation}\label{NLC:eq:1PK:frozen}
\widehat{\textbf{P}}^{(nc)}=\widehat{\textbf{T}}^{(nc)}\text{Cof}(\textbf{F}).
\end{equation}
The associated second Piola-Kirchhoff stress tensor is
\begin{equation}\label{NLC:eq:2PK:frozen}
\widehat{\textbf{S}}^{(nc)}=\textbf{F}^{-1}\widehat{\textbf{P}}^{(nc)}.
\end{equation}

\setcounter{equation}{0}
\renewcommand{\theequation}{B.\arabic{equation}}
\section{Detailed calculations for the energy components}\label{NLC:sec:append:bcs}
We provide in this appendix some detailed calculations for our derivation of the plate equations and their boundary conditions. These are standard and are included here for convenience. Similar calculations can be found in more detail in \cite{Landau:1986:LL}, for example.

First, we compute the different terms appearing in the stretching energy \eqref{NLC:eq:Estretch}. To lowest-order, the incompressibility condition is given by
\begin{equation}
\begin{split}
0&=\det\left(2\textbf{E}^{(0)}+1\right)-1\\
&=\left(2E^{(0)}_{33}+1\right)\left[\left(2E^{(0)}_{11}+1\right)\left(2E^{(0)}_{22}+1\right)-4\left(E^{(0)}_{12}\right)^2\right]-1\\
&= 2\left(E^{(0)}_{11}+E^{(0)}_{22}+E^{(0)}_{33}\right) +4\left[E^{(0)}_{11}E^{(0)}_{22}+E^{(0)}_{22}E^{(0)}_{33}+E^{(0)}_{33}E^{(0)}_{11}-\left(E^{(0)}_{12}\right)^2\right]+\mathcal{O}(\xi^6/L^6)\\
&= 2\text{tr}\ \textbf{E}^{(0)}+\mathcal{O}(\xi^4/L^4).
\end{split}
\end{equation}
It follows that $E^{(0)}_{33}=-E^{(0)}_{11}-E^{(0)}_{22}+\mathcal{O}(\xi^4/L^4)$. Similarly, $e^{(0)}_{33}=-e^{(0)}_{11}-e^{(0)}_{22}+\mathcal{O}(\xi^4/L^4)$.

Then,
\begin{equation}\label{NLC:eq:I10}
I_{1}^{(0)}=\text{tr}\left(2\textbf{E}^{(0)}+1\right)
=2\text{tr}\ \textbf{E}^{(0)}+3
=3+\mathcal{O}(\xi^4/L^4)
\end{equation}
and, with $p^{(0)}$ defined in \eqref{NLC:eq:p0}, we have
\begin{equation}
\begin{split}
-p^{(0)}\left(\mathcal{D}^{(0)}-1\right)&= -p^{(0)}\left[\det\left(2\textbf{E}^{(0)}+1\right)-1\right]\\
&= 4\mu\left[\left(E^{(0)}_{11}\right)^2+\left(E^{(0)}_{22}\right)^2+\left(E^{(0)}_{12}\right)^2+E^{(0)}_{11}E^{(0)}_{22}\right]+\mathcal{O}(\xi^6/L^6)\\
&= S^{(0)}_{11}E^{(0)}_{11}+S^{(0)}_{22}E^{(0)}_{22}+2S^{(0)}_{12}E^{(0)}_{12}+\mathcal{O}(\xi^6/L^6),
\end{split}
\end{equation}
where $E^{(0)}_{\alpha\beta}$ and $S^{(0)}_{\alpha\beta}$ are given by \eqref{NLC:eq:E0} and \eqref{NLC:eq:S0}, respectively. Therefore, 
\begin{equation}
E_{stretch}= H\int_{-L_{2}/2}^{L_{2}/2}\int_{-L_{1}/2}^{L_{1}/2}S^{(0)}_{\alpha\beta}E^{(0)}_{\alpha\beta}dX_{1}dX_{2},
\end{equation}
from which its first variation given in \eqref{NLC:eq:varEstretch} follows.
		
Second, we consider the bending energy \eqref{NLC:eq:Ebend}. We have 
\begin{equation}\label{NLC:eq:I12}
\begin{split}
I_{1}^{(2)}&=\left(\frac{\partial^2\xi}{\partial X_{1}^2}\right)^2
+\left(\frac{\partial^2\xi}{\partial X_{2}^2}\right)^2
+2\left(\frac{\partial^2\xi}{\partial X_{1}\partial X_{2}}\right)^2+\mathcal{O}(\xi^3/L^3)\\
&=\left(\Delta\xi\right)^2-2[\xi,\xi]+ \mathcal{O}(\xi^3/L^3)
\end{split}
\end{equation}
and
\begin{equation}
\mathcal{D}^{(2)}=[\xi,\xi]+\mathcal{O}(\xi^4/L^4),
\end{equation}
where $[\xi,\xi]$ is given by \eqref{NLC:eq:gcurvature}. 

The components of the associated Green-Lagrange strain tensor are
\begin{equation}\label{NLC:eq:E2} 
E^{(2)}_{\alpha\beta}=-\left(\frac{\partial^2\xi}{\partial X_{\alpha}\partial X_{\beta}}\right)^2, \qquad \alpha,\beta=1,2,
\end{equation}
and the corresponding second Piola-Kirchhoff stress has the components
\begin{equation}\label{NLC:eq:S2} 
S^{(2)}_{\alpha\beta}=2\mu E^{(2)}_{\alpha\beta}, \qquad \alpha,\beta=1,2.
\end{equation}
Therefore, 
\begin{equation}
\begin{split}
E_{bend}&=\frac{H^3}{12}\int_{-L_{2}/2}^{L_{2}/2}\int_{-L_{1}/2}^{L_{1}/2}S^{(2)}_{\alpha\beta}E^{(2)}_{\alpha\beta}dX_{1}dX_{2}\\
&=\frac{\mu H^3}{6}\int_{-L_{2}/2}^{L_{2}/2}\int_{-L_{1}/2}^{L_{1}/2} I_{1}^{(2)}dX_{1}dX_{2}.
\end{split}
\end{equation}

For the  variation given by \eqref{NLC:eq:varEbend}, we require the variational derivative
\begin{equation}
\frac{1}{2}\delta\left(\Delta\xi\right)^2
=\Delta\xi\Delta\delta\xi\\
=\nabla\cdot\left[\Delta\xi\nabla\delta\xi\right]-\nabla\cdot\left[\delta\xi\nabla\left(\Delta\xi\right)\right]+\delta\xi\Delta^2\xi,
\end{equation}
as well as the variational derivative of the Gaussian curvature,
\begin{equation}\label{NLC:eq:varGaussian}
\begin{split}
\delta [\xi,\xi]
&=\frac{\partial^2\xi}{\partial X_{1}^2}\frac{\partial^2\delta\xi}{\partial^2 X_{2}}
+\frac{\partial^2\delta\xi}{\partial X_{1}^2}\frac{\partial^2\xi}{\partial^2 X_{2}}
-2\frac{\partial^2\xi}{\partial X_{1}\partial X_{2}}\frac{\partial^2\delta\xi}{\partial X_{1}\partial X_{2}}\\
&=\frac{\partial}{\partial X_{1}}\left(\frac{\partial\delta\xi}{\partial X_{1}}\frac{\partial^2\xi}{\partial X_{2}^2}-\frac{\partial\delta\xi}{\partial X_{2}}\frac{\partial^2\xi}{\partial X_{1}\partial X_{2}}\right)
+\frac{\partial}{\partial X_{2}}\left(\frac{\partial\delta\xi}{\partial X_{2}}\frac{\partial^2\xi}{\partial X_{1}^2}-\frac{\partial\delta\xi}{\partial X_{1}}\frac{\partial^2\xi}{\partial X_{1}\partial X_{2}}\right).
\end{split}
\end{equation}
We define
\begin{equation}\label{NLC:eq:C1}
\mathcal{C}_{1}=\iint_{\mathcal{D}}\nabla\cdot\left(\Delta\xi\nabla\delta\xi\right)dA=\oint_{\mathcal{C}}\Delta\xi\left(\widetilde{\textbf{n}}\cdot\nabla\delta\xi\right)dL,
\end{equation}
\begin{equation}\label{NLC:eq:C2}
\mathcal{C}_{2}=\iint_{\mathcal{D}}\nabla\cdot\left[\delta\xi\nabla\left(\Delta\xi\right)\right]dA
=\oint_{\mathcal{C}}\delta\xi\left[\widetilde{\textbf{n}}\cdot\nabla\left(\Delta\xi\right)\right]dL,
\end{equation}
and
\begin{equation}\label{NLC:eq:C3}
\begin{split}
\mathcal{C}_{3}
=&\iint_{\mathcal{D}}\delta [\xi,\xi]dA\\
=&\oint_{\mathcal{C}}\left(\widetilde{\textbf{n}}\cdot\nabla\delta\xi\right)\left(
\widetilde{n}_{2}^2\frac{\partial^2\xi}{\partial X_{1}^2}
+\widetilde{n}_{1}^2\frac{\partial^2\xi}{\partial X_{2}^2}
-2\widetilde{n}_{1}\widetilde{n}_{2}\frac{\partial^2\xi}{\partial X_{1}\partial X_{2}}\right)dL\\
&+\oint_{\mathcal{C}}\delta\xi\widetilde{\textbf{t}}\cdot\nabla\left[
\widetilde{n}_{1}\widetilde{n}_{2}\left(\frac{\partial^2\xi}{\partial X_{2}^2}-\frac{\partial^2\xi}{\partial X_{1}^2}\right)+\left(\widetilde{n}_{1}^2-\widetilde{n}_{2}^2\right)\frac{\partial^2\xi}{\partial X_{1}\partial X_{2}}
\right]dL,
\end{split}
\end{equation}
where $\widetilde{\textbf{n}}=[\widetilde{n}_{1},\widetilde{n}_{2}]^{T}$ and $\widetilde{\textbf{t}}=[-\widetilde{n}_{2},\widetilde{n}_{1}]^{T}$ are the outward unit normal and the tangent vector to the boundary, respectively.
	
The boundary contribution in the expression of the small variation of bending energy \eqref{NLC:eq:varEbend} is then
\begin{equation}\label{NLC:eq:C}
\mathcal{C}=\frac{\mu H^3}{3}\left(\mathcal{C}_{1}-\mathcal{C}_{2}-\mathcal{C}_{3}\right).
\end{equation}

In particular, when the `edge' of the plate is free from traction, the variations $\delta\xi$ and $\textbf{n}\cdot\nabla\delta\xi$ are arbitrary, and their coefficients in the contour integrals are equal to zero. In this case, the boundary conditions for out-of-plane deformations are
\begin{equation}\label{NLC:eq:bc3:app}
\Delta\xi-\left(\widetilde{n}_{2}^2\frac{\partial^2\xi}{\partial X_{1}^2}
+\widetilde{n}_{1}^2\frac{\partial^2\xi}{\partial X_{2}^2}
-2\widetilde{n}_{1}\widetilde{n}_{2}\frac{\partial^2\xi}{\partial X_{1}\partial X_{2}}\right)=0,
\end{equation}
\begin{equation}\label{NLC:eq:bc4:app}
\widetilde{\textbf{n}}\cdot\nabla\left(\Delta\xi\right)+\widetilde{\textbf{t}}\cdot\nabla\left[
\widetilde{n}_{1}\widetilde{n}_{2}\left(\frac{\partial^2\xi}{\partial X_{2}^2}-\frac{\partial^2\xi}{\partial X_{1}^2}\right)
+\left(\widetilde{n}_{1}^2-\widetilde{n}_{2}^2\right)\frac{\partial^2\xi}{\partial X_{1}\partial X_{2}}
\right]=0.
\end{equation}

\setcounter{equation}{0}
\renewcommand{\theequation}{C.\arabic{equation}}
\section{An inflated nematic ring}\label{NLC:sec:append:ring}

Here, we derive the in-plane solution \eqref{NLC:eq:S0:sol}-\eqref{NLC:eq:u:sol} for the nematic ring under the boundary conditions specified in Section~\ref{NLC:sec:ring2}. We first consider a circular ring of neo-Hookean material, with the strain-energy function described by \eqref{NLC:eq:W:NH}. This ring occupies the reference configuration in cylindrical polar coordinates $(R,\Theta,Z)\in(R_{0},1)\times[0,2\pi)\times(0,H)$, where $R_{0},H>0$ are constants, and deforms in a plane perpendicular to its axis (see also \cite[p.~388]{goriely17}). In the current configuration, we have
\begin{equation}\label{NLC:eq:inflation:ring}
r=\frac{f(R)}{1+g_{1}},\qquad \theta=\frac{\Theta}{1+g_{2}},\qquad z=Z,
\end{equation}
where $f=f(R)$  is a function of $R$ to be determined. For this finite deformation, the Cauchy-Green tensor is equal to
\begin{equation}
\textbf{B}=\mathrm{diag}\left(\frac{\left(f'\right)^2}{\left(1+g_{1}\right)^{2}}, \frac{r^2}{R^2\left(1+g_{2}\right)^{2}},1\right),
\end{equation}
where $f'$ is the first derivative of $f$ with respect to $R$, such that the incompressibility condition $\det\textbf{B}=\left(f'/J\right)^2\left(r^2/R^2\right)=1$ is satisfied, and $J=\det\textbf{G}= \left(1+g_{1}\right)\left(1+g_{2}\right)$.

For the nematic ring described by the generalised Hooke's law, if the in-plane displacements and stresses only depend on $R$, then the Cauchy-Green tensor takes the approximate form \eqref{NLC:eq:B:app}, i.e., 
\begin{equation}\label{NLC:eq:B2B1:ring}
B_{22}\approx 1-2g_{2}, \qquad B_{11}=\frac{1}{B_{22}}\approx1+2g_{2}=1-2\left(g_{1}+1-J\right),\qquad B_{12}=0,
\end{equation}
where $J\approx 1+g_{1}+g_{2}$. The associated strain components given by \eqref{NLC:eq:e0} are
\begin{equation}\label{NLC:eq:strain:ring}
e_{11}=g_{2}=-g_{1}-1+J,\qquad e_{22}=-g_{2},\qquad e_{12}=0.
\end{equation}
It follows that the in-plane displacements are equal to
\begin{equation}\label{NLC:eq:u12:ring}
u_{1}=\left(R-R_{0}\right)\left(g_{1}+g_{2}\right),\qquad u_{2}=0,
\end{equation}
and the corresponding stresses are
\begin{equation}\label{NLC:eq:S0:ring}
S^{(0)}_{11}=2\mu g_{2},\qquad S^{(0)}_{22}=-2\mu g_{2},\qquad S^{(0)}_{12}=0.
\end{equation}
Thus, assuming $S^{(0)}_{11}=2\mu s_{0}$ at the outer surface implies $s_{0}=g_{2}$.


\end{document}